\documentclass[aps,prl,showpacs,twocolumn,superscriptaddress,floatfix]{revtex4-2}
\usepackage[utf8]{inputenc}
\newcommand{\figurescale}{1}
\usepackage{verbatim}
\usepackage{amsmath}
\usepackage{mathtools}
\usepackage{array}

\DeclarePairedDelimiterX\braket[2]{\langle}{\rangle}{#1 \delimsize\vert #2}

\usepackage[separate-uncertainty=true]{siunitx}
\DeclareSIUnit{\rpm}{rpm}

\usepackage[colorlinks=true,urlcolor=blue,linkcolor=blue,citecolor=blue]{hyperref}
\usepackage[T1]{fontenc}
\usepackage{placeins}
\usepackage{nicefrac}
\usepackage{pdfcomment}
\usepackage[usenames,dvipsnames]{xcolor}
\usepackage{braket}
\usepackage[normalem]{ulem}

\usepackage{graphicx}

\usepackage{lineno}

\begin{document}

\title{Defect complexes in CrSBr revealed through electron microscopy and deep learning}
%
%
\author{Mads~Weile}\email{weile.a.mads@gmail.com}
\affiliation{Department of Materials Science and Engineering, Massachusetts Institute of Technology, Cambridge, Massachusetts 02139, USA}
\affiliation{Center for Visualizing Catalytic Processes (VISION), Department of Physics, Technical University of Denmark, 2800 Kongens Lyngby, Denmark}
\author{Sergii~Grytsiuk}
\affiliation{Institute for Molecules and Materials, Radboud University, Heijendaalseweg 135, 6525AJ Nijmegen, The Netherlands}
\author{Aubrey~Penn}
\affiliation{MIT.nano, Massachusetts Institute of Technology, Cambridge, Massachusetts 02139, USA}
\author{Daniel~G.~Chica}
\affiliation{Department of Chemistry, Columbia University, New York 10027, United States}
\author{Xavier~Roy}
\affiliation{Department of Chemistry, Columbia University, New York 10027, United States}
\author{Kseniia~Mosina}
\affiliation{Department of Inorganic Chemistry, University of Chemistry and Technology Prague, Technická 5, 166 28 Prague 6, Czech Republic}
\author{Zdenek~Sofer}
\affiliation{Department of Inorganic Chemistry, University of Chemistry and Technology Prague, Technická 5, 166 28 Prague 6, Czech Republic}
\author{Jakob~Schiøtz}
\affiliation{Department of Physics, Technical University of Denmark, DK-2800 Kgs., Lyngby, Denmark}
\author{Stig Helveg}
\affiliation{Center for Visualizing Catalytic Processes (VISION), Department of Physics, Technical University of Denmark, 2800 Kongens Lyngby, Denmark}
\author{Malte~Rösner}
\affiliation{Institute for Molecules and Materials, Radboud University, Heijendaalseweg 135, 6525AJ Nijmegen, The Netherlands}
\author{Frances~M.~Ross}\email{fmross@mit.edu}
\affiliation{Department of Materials Science and Engineering, Massachusetts Institute of Technology, Cambridge, Massachusetts 02139, USA}
\author{Julian~Klein}\email{jpklein@mit.edu}
\affiliation{Department of Materials Science and Engineering, Massachusetts Institute of Technology, Cambridge, Massachusetts 02139, USA}
%
%
%
%
\date{2024-12-18}
%
%
\begin{abstract}
\textbf{
Atomic defects underpin the properties of van der Waals materials, and their understanding is essential for advancing quantum and energy technologies. Scanning transmission electron microscopy is a powerful tool for defect identification in atomically thin materials, and extending it to multilayer and beam-sensitive materials would accelerate their exploration. Here we establish a comprehensive defect library in a bilayer of the magnetic quasi-1D semiconductor CrSBr by combining atomic-resolution imaging, deep learning, and ab-initio calculations. We apply a custom-developed machine learning work flow to detect, classify and average point vacancy defects. This classification enables us to uncover several distinct Cr interstitial defect complexes, combined Cr and Br vacancy defect complexes and lines of vacancy defects that extend over many unit cells. We show that their occurrence is in agreement with our computed structures and binding energy densities, reflecting the intriguing layer interlocked crystal structure of CrSBr. Our ab-initio calculations show that the interstitial defect complexes give rise to highly localized electronic states. These states are of particular interest due to the reduced electronic dimensionality and magnetic properties of CrSBr and are furthermore predicted to be optically active. Our results broaden the scope of defect studies in challenging materials and reveal new defect types in bilayer CrSBr that can be extrapolated to the bulk and to over 20 materials belonging to the same FeOCl structural family.
}
\end{abstract}
%
%
\maketitle
%
%
\section{Introduction.}

Atomic defects such as vacancies, dopants or interstitials in two-dimensional (2D) van der Waals (vdW) materials have been studied extensively due to their ability to enhance the electronic, optical, magnetic, and chemical properties of the host material~\cite{Montblanch.2023, linDefectEngineeringTwodimensional2016}. Atomic modifications, particularly through defect control, are essential for engineering quantum emitters~\cite{Montblanch.2023}, fabricating spintronic devices~\cite{Atatre.2018} or enhancing catalytic activity~\cite{voiryEnhancedCatalyticActivity2013}. While defects have been extensively studied in graphene~\cite{Kotakoski.2011}, semiconducting transition metal dichalcogenides~\cite{Komsa.2012,Zhou.2013,Dumcenco.2013} and hexagonal boron nitride~\cite{Meyer.2009}, achieving atomic-scale insight is more challenging in 2D magnets and other correlated materials due to poor air stability~\cite{Gong.2017,Shcherbakov2018,Deng.2018,Bonilla2018,Liu2020}. However, defects in these materials are particularly interesting considering their impact on the local magnetic and charge order as well as their potential role in quantum light emission.

An exciting host for defects is the vdW magnetic semiconductor CrSBr~\cite{Katscher.1966,Gser.1990,Telford.2020}. Bulk CrSBr exhibits a direct bandgap of $\sim \SI{2}{\electronvolt}$~\cite{Watson.2024,Bianchi.2023,Klein.2023} and displays in-plane ferromagnetic and out-of-plane antiferromagnetic order~\cite{Telford.2020,Lee.2020}. The crystal structure of CrSBr is highly anisotropic, as seen in the material's quasi-1D electronic band structure that dominates quasiparticle physics~\cite{Klein.2023} and charge transport~\cite{Wu.2022}. Signatures of finite conductivity in low-temperature magneto-transport measurements~\cite{Telford.2022,Wu.2022} and spectrally narrow and magneto-correlated photoluminescence emission~\cite{Klein.2022a} have been tentatively ascribed to originate from magnetic defects. To date, the defect types and quantities in CrSBr are not well characterized, and more detail is needed if we are to understand how defects alter magneto-electronic properties or magnetic order~\cite{Klein.2022a,Long.2023,Tschudin.2024}. The recently observed phenomenon of Cr atom mobility under exposure to high energy electrons~\cite{Klein.2022}, as well as the use of helium ion irradiation~\cite{Torres.2023} to modify magnetic order~\cite{Long.2023}, motivate deliberate control of properties through defect control. In this case too, knowledge of the (induced) defect structures is required to predict and design properties.

\begin{figure*}
    \scalebox{\figurescale}{\includegraphics[width=1\linewidth]{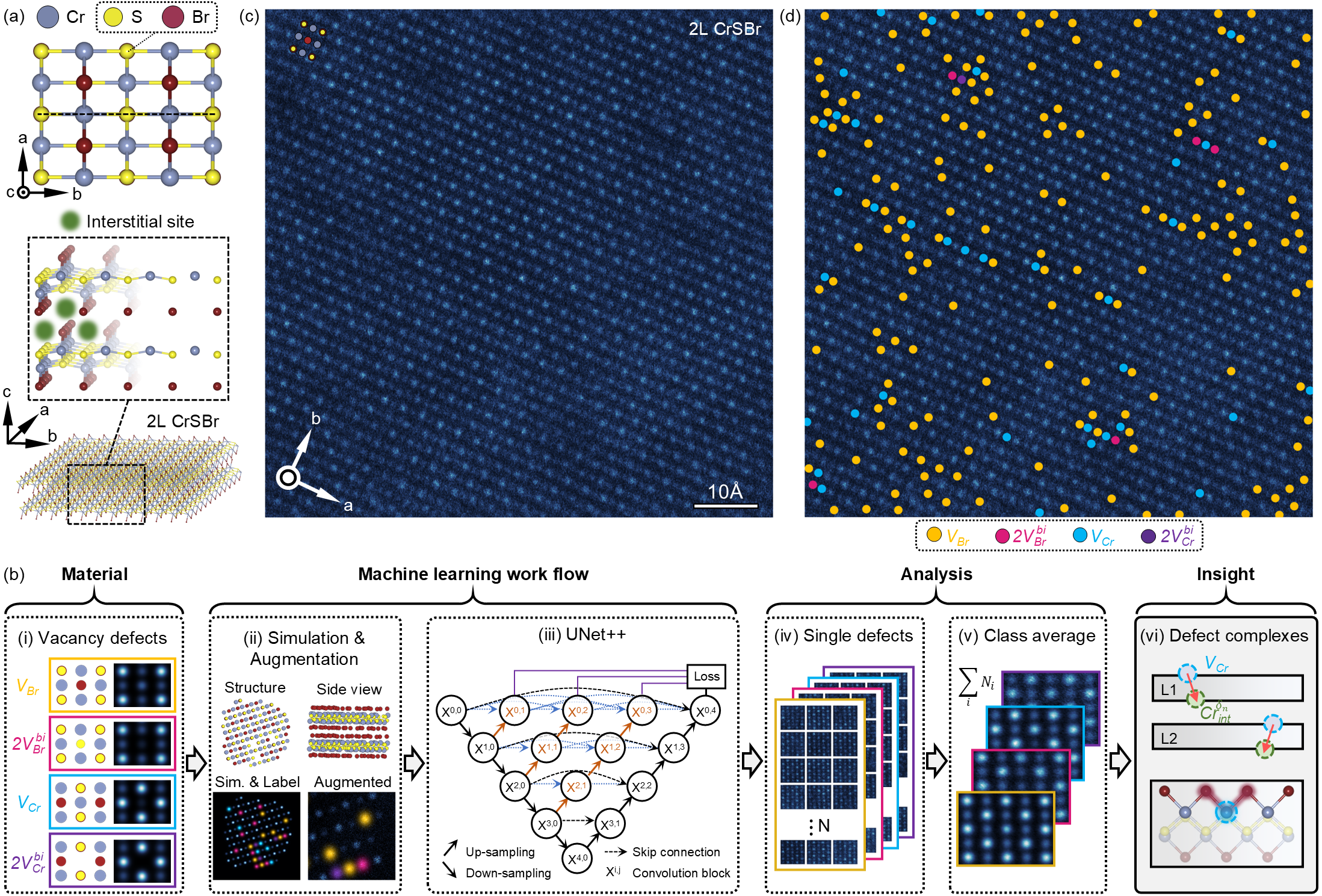}}
    \caption{\textbf{Detecting vacancy defects in 2L CrSBr utilizing a custom-built machine learning work flow.}
	(a) Crystal structure of 2L CrSBr and projected side view visualizing the atom stacking (S/Br on S/Br and Cr on Cr) in the b-c-plane. 
    (b) Step-by-step description of the work flow. (i) Four vacancy defects: V$_{Cr}$ (blue), V$_{Br}$ (purple), vertically stacked vacancies (in both layers) $V^{bi}_{2Cr}$ (orange), and $V^{bi}_{2Br}$ (pink) and their corresponding HAADF-STEM simulation. (ii) Image simulation and augmentation including vacancy defects and labelling. (iii) Neural network training using a UNet++ architecture~\cite{zhouUNetNestedUNet2018}. (iv) DCNN output from experimental images showing individual detected defects. (v) Class-averaging of classified vacancy defects to enhance SNR. (vi) Insights on defect complexes.
    (c) HAADF-STEM image of a 2L CrSBr using a beam current of $\SI{35}{\pico\ampere}$ and energy of $\SI{200}{\kilo\electronvolt}$. The total image dose is $9 \cdot 10^6 \SI{}{\mathrm{e^-}\per\nano\meter\squared}$.
    (d) Corresponding image with overlaid vacancy defects detected by DCNN. 
	}
    \label{fig:CrSBr_multilayer}
\end{figure*}

High-angle annular dark-field scanning transmission electron microscopy (HAADF-STEM) is a powerful tool that can assess atomic defects quantitatively in monolayer materials~\cite{Meyer.2009,Kotakoski.2011,Komsa.2012,Zhou.2013}. Low signal-to-noise-ratios (SNRs) for detecting individual atomic defects is a challenge in such studies, but for defect analysis in monolayers the use of machine learning techniques, particularly deep convolutional neural networks (DCNNs), has become a widespread strategy to mitigate low SNR~\cite{ziatdinovDeepLearningAtomically2017,madsenDeepLearningApproach2018,maksovDeepLearningAnalysis2019,Lee.2020a,Kalinin.2022}. In bilayer or multilayer materials the SNR of individual defects is even lower: enabling similarly quantitative studies in thicker crystals would accelerate material exploration. Bilayer (2L) CrSBr, as a precursor model system for bulk CrSBr, is particularly compelling, as defects break structural, electronic and magnetic symmetries to yield opportunities for optical emission~\cite{Klein.2022a}. Extending the application of DCNNs to thicker crystals would also be advantageous for the study of other beam-sensitive materials.

\begin{figure*}
    \scalebox{\figurescale}{\includegraphics[width=0.67\linewidth]{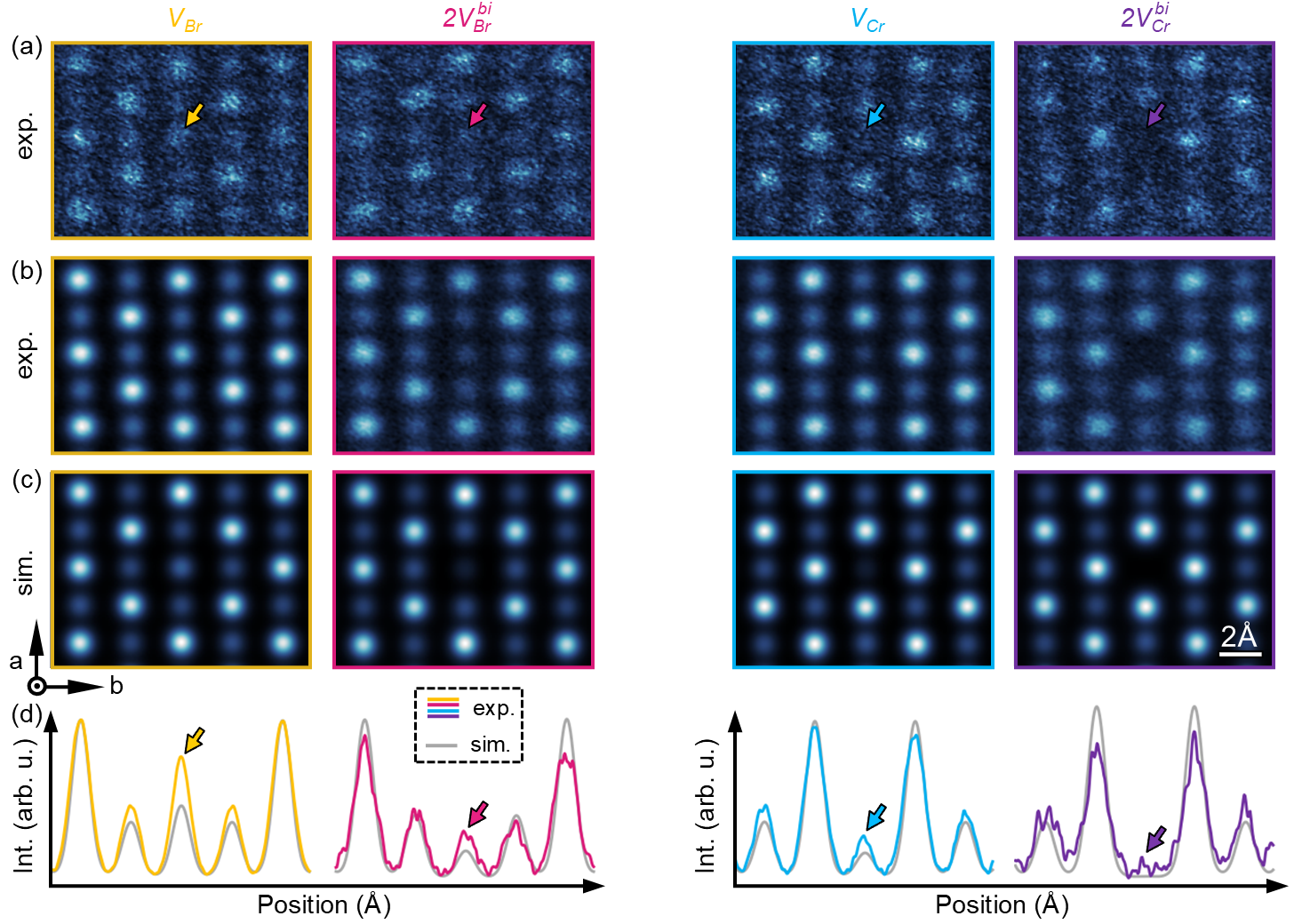}}
    \caption{\textbf{Single and vertically stacked vacancy defects in 2L CrSBr.} 
    (a) Single HAADF-STEM images of four types of vacancies ($V_{Br}$, $2V^{bi}_{Br}$, $V_{Cr}$ and $2V^{bi}_{Cr}$) detected by DCNN. The image size is $11.93 \SI{}{\angstrom}$ by $8.75 \SI{}{\angstrom}$. Arrows highlight the S/Br and Cr defect column, respectively.
    (b) Corresponding class-averaged images with a total of N = 1242, 64, 264 and 36 images per vacancy defect.
    (c) Corresponding simulated HAADF-STEM images from DFT relaxed crystal structures, performed using the experimental imaging parameters.
    (d) Comparison between experimental and simulated intensity line profiles from (b) and (c) along the b-direction.
    }
    \label{fig:class_average}
\end{figure*}

Here, we develop a STEM and deep learning approach that produces high quality images of atomic defects and defect complexes in 2L CrSBr and quantifies defect types and frequencies. We achieve this by applying a custom-developed machine learning work flow to detect and classify single point vacancy defects while also extending to vertically stacked vacancy defects. Our statistical approach of obtaining averaged images with improved SNR reveals a surprising group of defect complexes that were not initially anticipated in the analysis. One of these is a combined Cr vacancy and Cr interstitial defect complex that exists in different variants and symmetries owing to the vdW gap geometry and interstitial energy landscape in CrSBr; another is a combined Cr and Br vacancy complex that is characteristic for this compound, where Br atoms only share bonds with 2 Cr atoms. Moreover, from the statistical distribution of Br vacancies, we also reveal extended defect lines of different length with an occurrence suggesting correlated formation behavior. We show ab-initio DFT calculations of structures and binding energies that are in excellent agreement with our observations. In particular, the interstitial defect complexes exhibit strong electronic localization, and we suggest a range of quantum emission properties. Our study lays a foundation for utilizing the versatile group of defects that are evidently possible within CrSBr and can be extrapolated to many materials with the same FeOCl structural type. We conclude that the exceptional properties of CrSBr as a host for magneto-correlated and optically active defects, combined with its responsiveness to electron beam-induced structural transformation~\cite{Klein.2022}, suggest promising opportunities for controlled defect complex engineering.


\section{Methods}

CrSBr is a ternary compound in which each layer consists of a network of Cr and S atoms encapsulated by Br atoms (see Fig.~\ref{fig:CrSBr_multilayer}(a)). It shows several important structural features. First, S and Br atoms reside in their own atom column distinct from the Cr atom column (see Fig.~\ref{fig:CrSBr_multilayer}(a)). Second, Br atoms are exclusively bonded to Cr atoms forming a linear chain structure along the a-direction. Third, due to the AA layer stacking order~\cite{Beck.1990} with an interlayer distance of $\SI{7.965}{\angstrom}$~\cite{Gser.1990}, Br atom chains are displaced between adjacent layers to form interstitial sites between the layers and on the surface (see Fig.~\ref{fig:CrSBr_multilayer}(a)) in both atom columns. This crystal structure, of the FeOCl-type, is well known for intercalation~\cite{Sagua2001}. The three features described above are important in understanding the types of defects in CrSBr we identify in this work.

We begin by describing our custom-built work flow for detecting defects in 2L CrSBr (see Fig.~\ref{fig:CrSBr_multilayer}(b)). For our material, 2L CrSBr, we focus on four different defect types: single $V_{Br}$ and $V_{Cr}$ vacancies in either of the layers, and vertically stacked bi-vacancies $V^{bi}_{2Br}$ and $V^{bi}_{2Cr}$ in both layers. We focus on defects related to Cr and Br because they lie within our detection sensitivity. We find that detecting $V_S$ is challenging due to the low $Z$ contrast change, particularly in bilayers. This is in contrast to the assignment of Cr and Br related defects, as these exhibit distinguishable $Z$ contrast.

For the machine learning work flow and training of our DCNNs, we simulate 600 HAADF-STEM images of 30x30 Å$^2$ flakes of CrSBr. These have been subjected to a variety of different transformations and noise sources to generalize the network to understand experimental data (see SM). Using the simulated images, a ground-truth label is constructed by centering Gaussians at each corresponding defect location with a standard deviation of $\sigma=0.5 \mathrm{\textup{~\AA}}$. We find that realistic modelling of surface contamination is important to achieve robust results. We add fractal noise~\cite{eeveePerlinNoise2016} to the simulated HAADF-STEM images (see SM) to enhance the effectiveness of our DCNNs to detect defects in experimental images that exhibit this common type of contamination~\cite{Lin.2011,Dyck.2021}. We expect that other imaging modalities, such as differential phase contrast imaging or ptychography, should be equally applicable for this work flow particularly for light elements~\cite{Yang2016}.

\begin{figure*}
    \scalebox{\figurescale}{\includegraphics[width=1\linewidth]{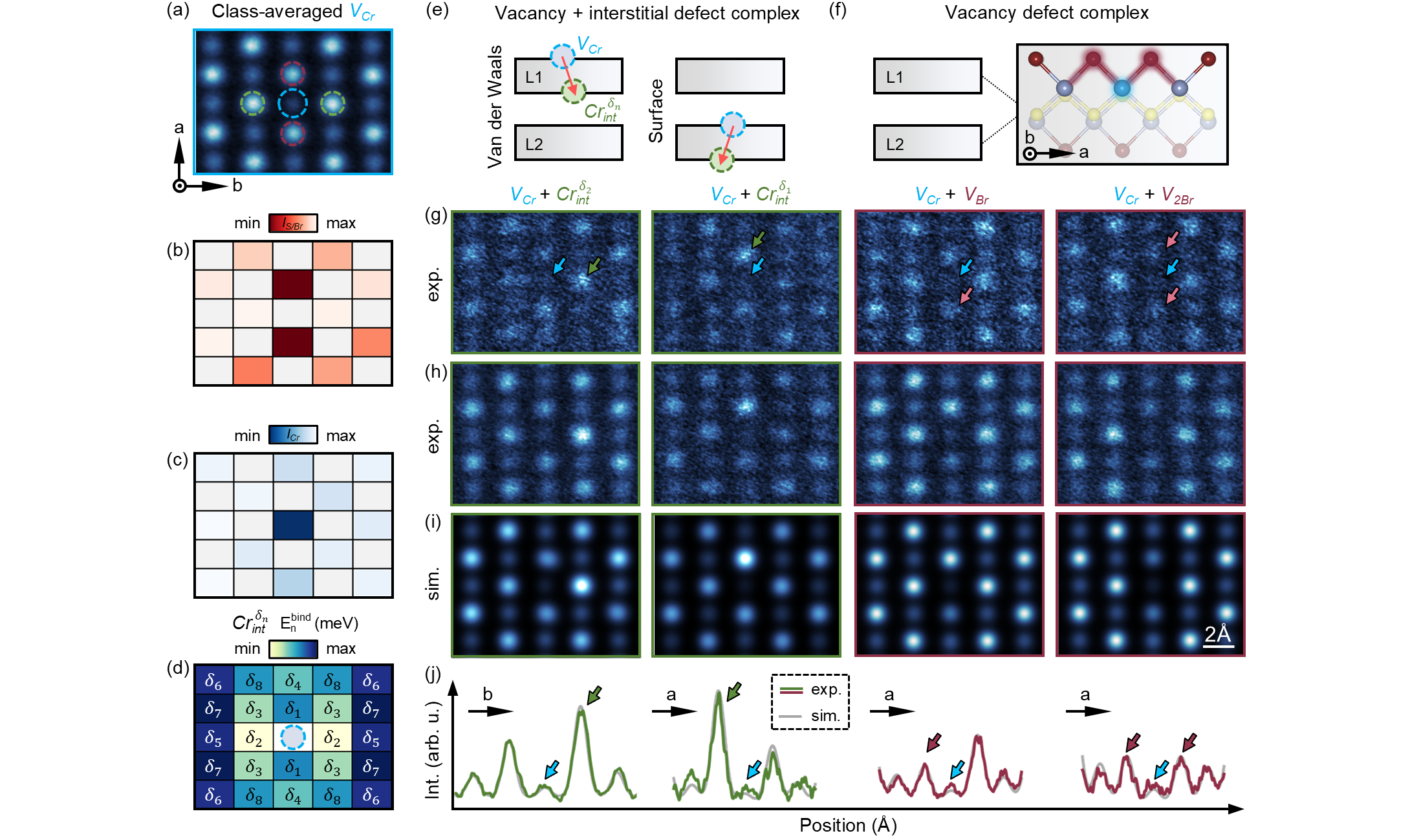}}
    \caption{\textbf{Interstitial and vacancy defect complexes in 2L CrSBr.}  
    (a) Class-averaged HAADF-STEM image of $V_{Cr}$ from Fig.~\ref{fig:class_average}(b). Nearest neighbor (NN) S/Br atom columns along the a- (red) and b-direction (green) are highlighted.
    (b), (c) Corresponding Voronoi intensity diagram of the S/Br and Cr atom column.
    (d) DFT calculated binding energy density of 1$^{st}$- and 2$^{nd}$-NN Cr interstitial defects.
    (e) Vacancy-interstitial defect complex with $Cr_i$ situated between layers (L1 and L2) in the van der Waals gap or at the surface. 
    (f) Vacancy defect complex due to exclusive Cr-Br bond.
    (g) Single experimental HAADF-STEM images of vacancy-interstitial defect complexes $V_{Cr}$+$Cr^{\delta 2}_{int}$ and $V_{Cr}$+$Cr^{\delta 1}_{int}$ as well as $V_{Cr}$+$V_{Br}$ and $V_{Cr}$+$V_{2Br}$ defect complexes.
    (h) Corresponding class-averaged images with a total of N = 11, 3, 12 and 6 images per defect type.
    (i) Simulated images from DFT relaxed crystal structures.    
    (j) Comparison between experimental and theoretical intensity line profiles from (h) and (i) along the b-direction or a-direction.
	}
    \label{fig:atlas_of_defects}
\end{figure*}

We employ two types of DCNNs, a defect network specialized in finding defects of interest and an atom-spotting network which is optimized for finding atomic positions (see SM)~\cite{ziatdinovBuildingExploringLibraries2019}. This allows the detection and subsequent alignment of defects with few-pm-precision. For both networks we use the UNet++ architecture~\cite{zhouUNetNestedUNet2018}. This architecture is a derivative of the classic UNet~\cite{UNET} and introduces additional pathways between the down- and up-sampling steps, effectively enabling an improved mapping across different length scales. The training process, parameters and labelling is discussed in detail in the SM and the code is also provided online~\cite{Weile2024}.

We collect sequences of 10 images of 2L CrSBr with a short dwell time ($\SI{500}{\nano\second}$/px) that are averaged using rigid registration~\cite{Savitzky.2018}, resulting in a total dose of $9 \cdot 10^6 \SI{} { \mathrm{ e^-}\per\nano\meter\squared}$ (see Fig.~\ref{fig:CrSBr_multilayer}(c)). Imaging is performed along the crystallographic c-axis. The image shows the characteristic two columns with higher brightness for the S/Br atom column, where $Z(2S+2Br)=102$, and lower brightness for the Cr atom column, where $Z(2Cr)=48$, following the relationship between HAADF detector intensity and atomic number $I \sim Z^{1.6-1.7}$~\cite{Krivanek.2010}. Moreover, we observe defective regions that we ascribe to the material transfer process that exposes the material to solvents and low heat, in addition to defects that may occur from ambient humidity and from the electron beam during imaging (see SM).

We train an ensemble of 4 DCNNs and apply them to experimental HAADF-STEM images to predict vacancy defects. Figure~\ref{fig:CrSBr_multilayer}(d) shows the HAADF-STEM image with overlaid predicted defects. Here, the most frequently detected defects are single vacancy defects, $V_{Br}$ and $V_{Cr}$. Less frequently, we also observe the vertically stacked vacancies ($2V^{bi}_{Br}$ and $2V^{bi}_{Cr}$), typically in regions that appear more defective. Generally, the $V_{Br}$ is the most abundant defect. The reduced stability of Br atoms is anticipated due to their position protruding from the layers. As halogens, Br atoms are particularly prone to surface chemical reactions, especially hydrolysis with water~\cite{Torres.2023}. Our expectation also aligns with calculated defect binding energy densities (see Tab.~\ref{tab:structures} and SM). A high concentration of $V_{Br}$ has also been observed in STM topographic images~\cite{Klein.2022a}. It is important to note that the densities we observe are substantially higher than the intrinsic defect concentration (see SM). We attribute this to the treatments required for sample preparation and the stability limitations of CrSBr (see SM).

\section{Results}

After establishing the material problem and introducing our approach, we now apply our custom-built work flow to detect vacancy defects in a set of 13 experimental HAADF-STEM images with a total area of $\SI{832}{\nano\meter\squared}$ (see SM). A representative single image of $V_{Br}$, $2V^{bi}_{Br}$, $V_{Cr}$, and $2V^{bi}_{Cr}$ and their corresponding class-averages are shown in Fig.~\ref{fig:class_average}(a) and (b), respectively. Obtaining statistical data on multiple images improves SNR, analogously to approaches used in cryo-electron microscopy~\cite{Cheng.2009}. Averaging is essential to analyze defects as the stability of 2L CrSBr only allows for low-dose imaging resulting in low SNR. We compare the experimental images with simulations (see Fig.~\ref{fig:class_average}(c)) that are based on DFT relaxed crystal structures. Line profiles taken along the b-direction show excellent quantitative and qualitative agreement between experiment and simulation (see Fig.~\ref{fig:class_average}(d)). The main disparity we observe is for the $V_{Br}$, where the central experimental intensity (arrowed) is higher than the simulated one. We attribute this discrepancy to the detection of false positives (identifying non-defects as defects). We note that the simulations should only be treated as a close approximation as reproducing experimental conditions is difficult~\cite{Krause.2013}. Under these challenging imaging conditions and low Z contrast, both false positives and false negatives (failing to detect actual defects) can occur. However, by employing class-averages we minimize their impact: false positives will result in less pronounced features in the averages, while false negatives do not contribute. Additionally, we further reduce false positives by applying a detection threshold of 0.95 for $V_{Br}$ (which has the lowest contrast change, compared to its pristine counterpart) and 0.90 for all other defects. As well as calculating class-averages, we also verify that our trained networks provide statistically significant outputs by performing two additional cross-validation approaches (see SM). 

The statistical approach using deep learning combined with the quantitative nature of collecting HAADF-STEM images equips us with the ability to spot not only the presence of vacancy sites, but potentially more intricate defects that either involve multi-site vacancy complexes or atoms residing in proximal atom columns as interstitials. The latter is particularly relevant for CrSBr considering the mobility of Cr atoms when exposed to high energy electrons~\cite{Klein.2022}.

With this in mind, we revisit atom column intensities of the 1$^{st}$- and 2$^{nd}$-nearest neighbor (NN) S/Br and Cr atom columns in the class-averaged $V_{Cr}$ (see Fig.~\ref{fig:atlas_of_defects}(a)) and create a corresponding Voronoi diagram by integrating a circular mask around each atom (see Fig.~\ref{fig:atlas_of_defects}(b) and (c))~\cite{E2013}. Because the class-average represents a statistical measurement across our 264 detected $V_{Cr}$ related defects, we can treat it as a probabilistic measure to observe other more complex defect types. We note that this only considers defects that are connected to the occurrence of a nearby $V_{Cr}$. We make two striking observations. First, there is reduced intensity in the 1$^{st}$-NN S/Br column along the a-direction, and second, there is increased intensity along the S/Br column along the b-direction. We interpret these intensities through the presence of two categories of defect complexes. One is a $V_{Cr}$ in combination with a Cr interstitial $Cr^{\delta n}_{int}$ with $\delta n$ as the specific position of the Cr interstitial atom (see Fig.~\ref{fig:atlas_of_defects}(e)), and the other is a vacancy complex consisting of $V_{Cr}$ and one or two $V_{Br}$ along the a-direction (see Fig.~\ref{fig:atlas_of_defects}(f)).

\begin{figure*}
    \scalebox{\figurescale}{\includegraphics[width=1\linewidth]{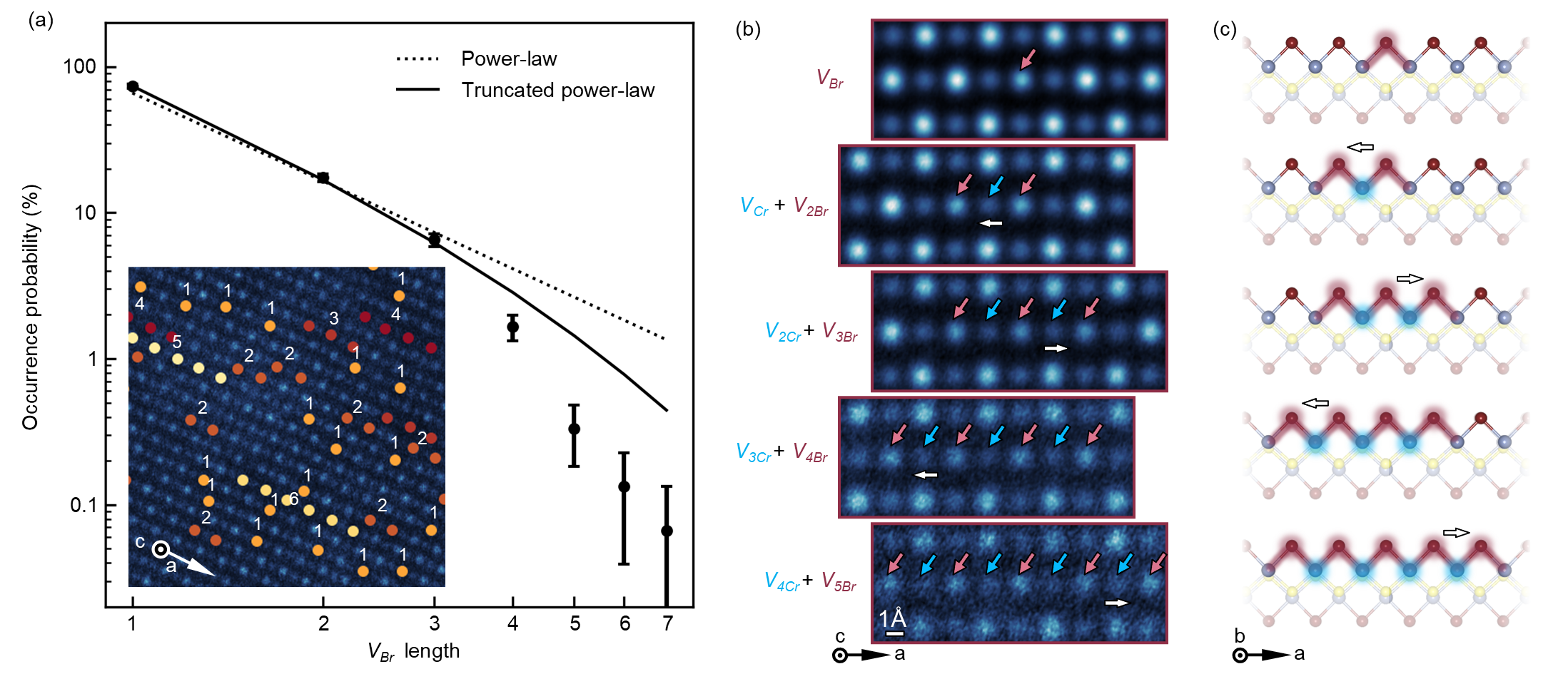}}
    \caption{\textbf{Extended 1D line defects in 2L CrSBr.}
    (a) Probability of defect chains of different length following a power-law dependence. Inset: HAADF-STEM image with detected Br vacancy defect lines of different length.
    (b) Class-averages with a total of N = 1109, 262, 98, 25, and 5 images per defect length type.
    (c) Illustration of defect chain growth along the a-direction.}
    \label{fig:line_defect}
\end{figure*}

We first discuss the class of Cr interstitial defect complexes. Formation probabilities can be estimated based on ab-initio DFT calculations. For this, we determine the binding energy density of eight Cr interstitial defects $V_{Cr}$+$Cr^{\delta n}_{int}$ with $n = 1-8$ representing the location of the interstitial (see Fig.~\ref{fig:atlas_of_defects}(d), Tab.~\ref{tab:structures} and SM). This is calculated by $E^\text{bind}_{Cr^{\delta n}_{int}} = \dfrac{E_N - \Big(E_{N-Cr^{\delta n}_{int}} + E_{Cr^{\delta n}_{int}}\Big)}{N_\text{cells}}$ which is defined as the energy to remove the $Cr^{\delta n}_{int}$ atom from the unit cell consisting of N-atoms. A lower $E^\text{bind}_{n}$ suggests a higher probability of forming a Cr interstitial defect complex. We find the $V_{Cr}$+$Cr^{\delta 2}_{int}$ to be energetically most favorable. Importantly, in 2L CrSBr, and more generally in multilayer CrSBr, we need to distinguish further between a Cr interstitial that is situated between vdW layers and one at the surface (see Fig.~\ref{fig:atlas_of_defects}(e)). As may be expected, interstitials located in the vdW gap are calculated to be more stable compared to interstitials situated at the surface (see Tab.~\ref{tab:structures} and SM).

Manually searching through images of all detected $V_{Cr}$ and locating increased intensity in 1$^{st}$- and 2$^{nd}$-NN S/Br and Cr atom columns, we identify several prominent types of Cr interstitial defect complexes (see Fig.~\ref{fig:atlas_of_defects}(g) and (h)). Most common is the $V_{Cr}$+$Cr^{\delta 2}_{int}$ (Cr in the S/Br column along the b-direction). Its relatively frequent occurrence is in excellent agreement with the lowest binding energy density $E^{bind}_{\delta 2} = \SI{105}{\milli\electronvolt}$ (see Fig.~\ref{fig:atlas_of_defects}(d)). Moreover, we also observe interstitial defects with the Cr atom residing in the closest S/Br atom column along the a-direction to form the $V_{Cr}$+$Cr^{\delta 1}_{int}$ complex ($E^{bind}_{\delta 1} = \SI{149}{\milli\electronvolt}$). It is unknown whether the observed interstitial is thermally or kinetically stable, however a match with DFT calculations of lowest energy motifs suggest that irrespective of the formation mechanism the observed features seem to be thermally equilibrated. Indeed, we also observe this type of relocation of Cr atoms during beam exposure in 2L CrSBr (see SM). We further note that it is unclear whether all observed interstitials are beam-induced or formed during sample fabrication or material synthesis. The class-averages show excellent agreement with simulated images of corresponding DFT relaxed crystal structures (see Fig.~\ref{fig:atlas_of_defects}(i)), as assessed by comparing line profiles along the b- and a-directions (see Fig.~\ref{fig:atlas_of_defects}(j)). We note that the different configurations and locations of the Cr interstitial (vdW and surface) create lattice distortions that result in qualitative differences in the simulated image intensities (see SM). While we have focused on the $V_{Cr}$+$Cr^{\delta 1}_{int}$ and $V_{Cr}$+$Cr^{\delta 2}_{int}$ interstitial defect complexes, other Cr interstitial defect complexes may also be formed under the electron beam.

We now discuss the vacancy complex consisting of $V_{Cr}$ with $V_{Br}$. Using the same approach as in searching for Cr interstitials, we manually assess the $V_{Cr}$ for an intensity reduction at the position indicated in Fig.~\ref{fig:atlas_of_defects}(a) as suggested from our Voronoi diagram (see Fig.~\ref{fig:atlas_of_defects}(b)). We do indeed identify the presence of $V_{Cr}$+$V_{Br}$ and $V_{Cr}$+$2V_{Br}$ vacancy defect complexes (see Fig.~\ref{fig:atlas_of_defects}(g) and (h)). The presence of such complexes is intuitive as the Br atoms are exclusively bonded to the Cr atoms (see Fig.~\ref{fig:atlas_of_defects}(f)) and the presence of $V_{Cr}$ lowers $E^\text{bind}$ of a Br atom from $\SI{350}{\milli\electronvolt}$ to $\SI{306}{\milli\electronvolt}$ for $V_{Br}$ or $\SI{293}{\milli\electronvolt}$ for $2V_{Br}$ (see Tab.~\ref{tab:structures} and SM). Line profiles from the class-averaged images obtained from the manual search are in excellent agreement with simulated images from DFT relaxed structures (see Fig.~\ref{fig:atlas_of_defects}(i) and (j)). Computationally searching through all $V_{Cr}$ that show $V_{Br}$ or $2V_{Br}$ in the 1$^{st}$-NN S/Br columns, we find that $V_{Cr}$ is concurrent with $V_{Br}$ $43\%$ of the time and $V_{2Br}$ with $19\%$. This suggests the frequent formation of $V_{Cr}$+$V_{Br}$ and $V_{Cr}$+$V_{2Br}$ defect complexes, most likely at the surface of the bilayer. We note that purely based on HAADF-STEM images we are not able to distinguish whether the vacancy is the Br atom site at the surface or the one close to the vdW gap.

\begin{figure*}
    \scalebox{\figurescale}{\includegraphics[width=1\linewidth]{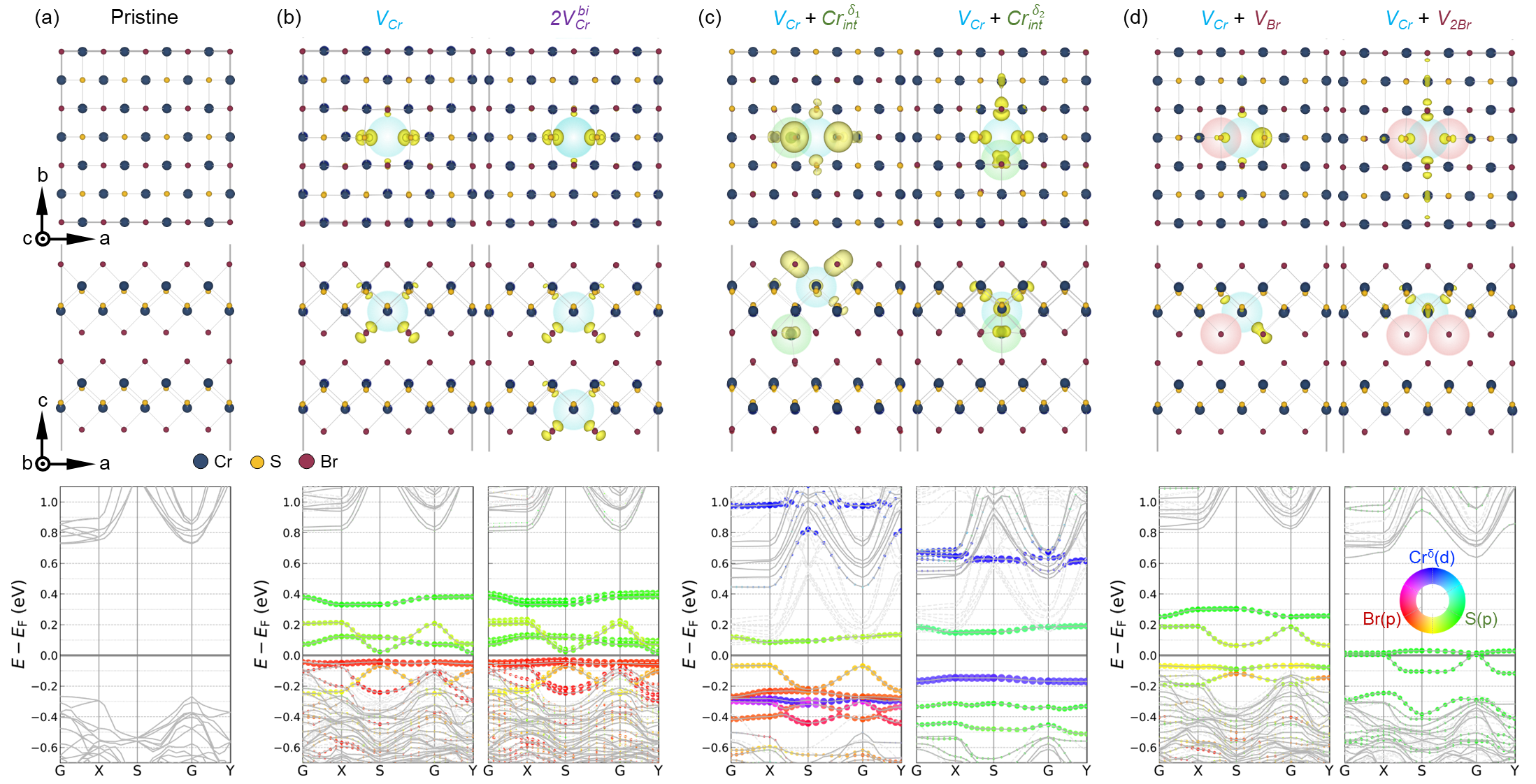}}
    \caption{\textbf{Electronic structure of single defects and defect complexes in 2L CrSBr.}
    Upper panels show the DFT+U relaxed crystal structures and defect wave functions for 
    (a) the pristine 2L CrSBr $4 \times 3$ supercell, 
    (b) single Cr vacancies ($V_{Cr}$ and $V^{bi}_{2Cr}$), 
    (c) interstitial defects ($V_{Cr}$+$Cr^{\delta 1}_{int}$, $V_{Cr}$+$Cr^{\delta 2}_{int}$), and 
    (d) two vacancy complexes ($V_{Cr}$+$V_{Br}$ and $V_{Cr}$+$V_{2Br}$). The charge densities associated with defects are shown in yellow, while Cr and Br vacancies and Cr interstitial atoms are indicated by light blue, dark red and green spheres, respectively.
    Lower panels show the electronic band structures for the same configurations. Colored bands represent defect-induced majority orbital wave function admixtures from Cr \textit{d}-orbitals and \textit{p}-orbitals of S and Br in the defect vicinity. Solid lines in gray denote the majority spin channel while dashed lines indicate the minority spin channel.
    }
    \label{fig:theory}
\end{figure*}


Considering the volatility of surface Br atoms and their arrangement in chains along the a-direction, we now study their statistical distribution more closely to assess potential correlations based on their occurrence and spatial arrangement. To this end, we analyze and quantify both individual $V_{Br}$ and extended $V_{Br}$ defect chains along the a-direction (see Fig.~\ref{fig:line_defect}(a) and SM). Interestingly, we find that the statistical distribution of Br vacancy defect lengths is best captured by a power-law distribution (see Fig.~\ref{fig:line_defect}(a) and SM), suggesting that $V_{Br}$ exhibit correlated, extended growth patterns. We observe that deviations from the expected power-law behavior may result from false positives or false negatives in the detection of Br defects. Specifically, false positives can lead to an increased number of detected individual $V_{Br}$ defects, while false negatives can reduce the count of longer extended defect lines.

While a Br vacancy can initiate the formation of defect chains, a Cr vacancy is an even more likely nucleation site, as it significantly lowers the defect formation energy of the two neighboring Br atoms to which it is bonded to. Indeed, in our class-averaged images of different defect lengths (see Fig.~\ref{fig:line_defect}(b)), we not only see $V_{Br}$ but also $V_{Cr}$ participate in forming extended 1D defect lines. The presence of either vacancy reduces the formation energy of adjacent defects. For example, once a $V_{Cr}$ forms, the subsequent defect formation energy for an adjacent $V_{Br}$ decreases, which facilitates the formation of extended line defects (see Fig.~\ref{fig:line_defect}(c) and SM). As this mechanism repeats, long defective chains composed of missing Br and Cr atoms can develop along the a-direction. Based on our electronic structure calculations (see SM), we anticipate the formation of various types of line defects. Among these, the pure $V_{Br}$ defects, along with $V_{Br}$ in conjunction with fewer $V_{Cr}$ vacancies, are expected to represent the most energetically favorable configurations.

Several factors likely contribute to the correlated formation and extension of these defect lines. The relatively low formation energy of $V_{Br}$ makes Br atoms particularly volatile. Mechanisms that can drive the growth of defect lines include surface reconstruction, strain relaxation, or external chemical influences, such as hydrolysis from H$_2$0 or oxidation from O$_2$~\cite{Torres.2023}. Such reactions are particularly dominant on surface layers.

The identification and quantification of Br vacancy defects, defect complexes and extended line defects are relevant for interpreting transport measurements on CrSBr~\cite{Wu.2022,Telford.2022}. Recent conductivity measurements reported an extreme anisotropy between the a- and b-direction with a strong gate voltage dependence~\cite{Wu.2022}. Despite the characteristic flat conduction band along the a-direction and a highly dispersive band along the b-direction with an effective electron mass ratio of $\sim 50$~\cite{Klein.2023}, the highest anisotropy is observed for hole doping but is suppressed for electron doping~\cite{Wu.2022}. This unexpected response suggests that defects may be important for interpreting the transport measurements. Our calculations show that the most abundant type of defect, the $V_{Br}$, induces a shallow donor-like states close to the conduction band minimum (see SM), effectively n-doping CrSBr, as also observed in transport experiments~\cite{Wu.2022,Telford.2022}. Moreover, from our calculations of smaller supercells, we also see defect-line induced dispersions along the a- and b-direction that could potentially promote hopping transport. In our study, we not only verify a high concentration of $V_{Br}$ but more strikingly also observe the formation of 1D defect lines ($V_{(n)Br}$ and $V_{(n)Br}$+$V_{(n-1)Cr}$) with up to $n = 7$ Br atoms extending for more than $\SI{2}{\nano\meter}$. To study their effect on the electronic band structure, and therefore transport properties, we calculate infinite vacancy lines along the a-direction (see SM). These calculations show dispersion and metallic behavior along the a- but also the b-direction. This suggests that defect lines not only contribute free electrons, increasing the intrinsic carrier concentration, but can likely also contribute to the gate voltage dependent conductivity anisotropy observed along the a- and b-direction.

These results imply that surface vacancy engineering and chemical manipulation can provide a means to control electronic transport properties in this material system, potentially enabling tunable anisotropic transport properties through deliberate defect engineering. This becomes particularly important for mono- and bilayer CrSBr flakes due to their high surface-to-volume ratio, but is also important for bulk material where it is expected that defects in surface layers contribute disproportionately to bulk device transport due to their higher density of interface states.


We can now discuss the magneto-electronic footprints of the defects that are most frequently observed experimentally. We calculate the electronic structure by means of spin-resolved DFT+U calculations for $4 \times 3$ supercells of 2L CrSBr. We are particularly interested in the coupled electronic and magnetic degrees of freedom of the defects, as CrSBr exhibits strongly magneto-correlated physics~\cite{Telford.2020,Lee.2020,Wilson.2021,Klein.2022a,Klein.2023}. Moreover, we are also interested in the magnetic coupling of the Cr interstitials as they are expected to affect local magnetic intra- and interlayer ordering~\cite{Klein.2022,Klein.2022a,Long.2023}.

We calculate the electronic band structure of pristine (see Fig.~\ref{fig:theory}(a)) and defective 2L CrSBr supercells (see Fig.~\ref{fig:theory}(b)-(d)) alongside their real-space defect charge densities. For the defective band structures we consider the Cr vacancy defects $V_{Cr}$ and $2V^{bi}_{Cr}$ (see Fig.~\ref{fig:theory}(b)), the Cr interstitial vacancy complexes $V_{Cr}$+$Cr^{\delta 2}_{int}$ and $V_{Cr}$+$Cr^{\delta 1}_{int}$(see Fig.~\ref{fig:theory}(c)), and the Cr and Br vacancy complexes $V_{Cr}$+$V_{Br}$ and $V_{Cr}$+$V_{2Br}$ (see Fig.~\ref{fig:theory}(d) and SM). 

We note that in the LDA+U approximation the band gap of the pristine 2L CrSBr is underestimated. Otherwise, we find in Fig.~\ref{fig:theory}(a) the expected characteristics such as highly anisotropic dispersions in the lowest conduction bands as well as the antiferromagnetic alignment between the ferromagnetic single layers, which combined with the inversion symmetry of the 2L system results in a fully spin-degenerate electronic structure. In all calculated defect structures we find multiple weakly dispersive in-gap states, indicating the highly localized character of these impurity states in real space (which still hybridize here due to periodic boundary conditions and the finite size of the supercells).

In general, as illustrated in Fig.~\ref{fig:theory}, the atoms surrounding the defects predominantly contribute to the impurity states near the Fermi level, exhibiting distinct orbital characters for various defects. Although the $p$-orbital of sulfur atoms is present in all cases, the $d$-orbital significantly contributes only in the presence of Cr displacements ($V_{Cr}$ + $Cr^{\delta}_\text{int}$ defects). In structures with pure Cr vacancy defects, there is a notable contribution from the $p$-orbital of Br atoms; however, this contribution nearly disappears when two Br atoms adjacent to the Cr vacancy are removed. 
Additionally, the contribution from the Br $p$-orbitals is highly dependent on the displacement direction of the Cr atoms and the location of the resulting Cr vacancy, peaking for the $\delta_1$ displacement and nearly vanishing for the $\delta_2$ displacement.

\begin{figure*}
    \scalebox{\figurescale}{\includegraphics[width=1\linewidth]{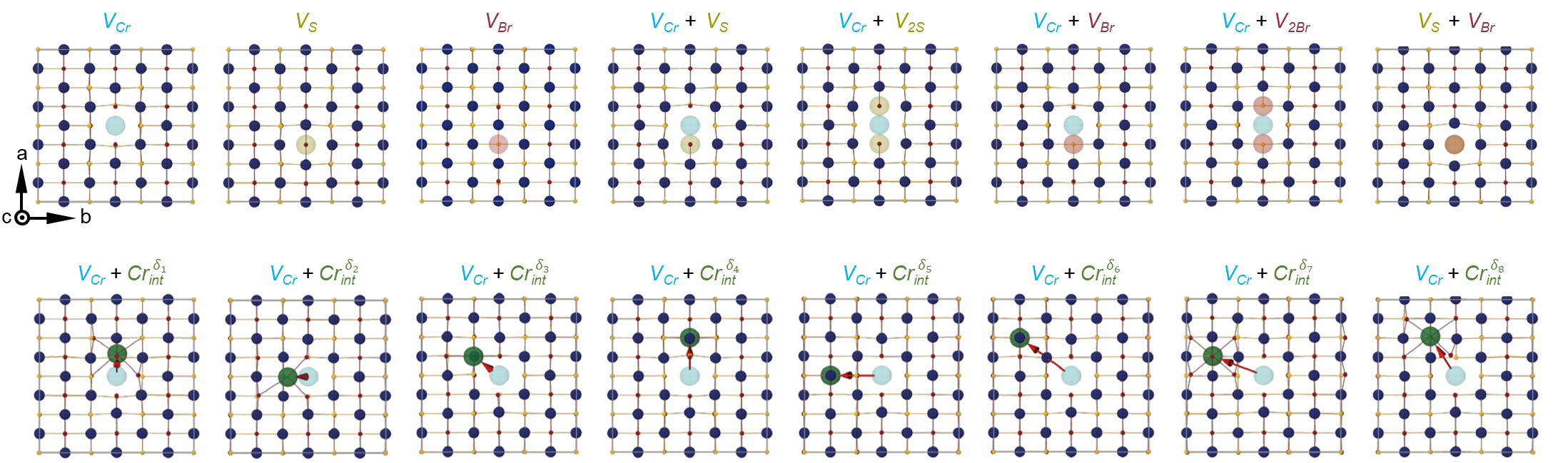}}
    \caption{\textbf{Library of defects and defect complexes in 2L CrSBr.}
    DFT-relaxed crystal structures depicting the top-down view of vacancy defects ($V_{Cr}$, $V_{S}$ and $V_{Br}$), vacancy defect complexes ($V_{Cr}$+$V_{S}$, $V_{Cr}$+$V_{2S}$, $V_{Cr}$+$V_{Br}$, $V_{Cr}$+$V_{2Br}$ and $V_{S}$+$V_{Br}$) and vacancy interstitial complexes ($V_{Cr}$+$Cr^{\delta n}_{int}$).
	}
    \label{fig:theory_structures}
\end{figure*}

In the case of the single Cr vacancy $V_{Cr}$ depicted in Fig.~\ref{fig:theory}(b), we find various defect in-gap states around the Fermi level with partially vanishing dispersion. Most importantly, we see that the impurity states inherit the spin state of the monolayer in which the defect is situated. Consequently, the $2V^{bi}_{Cr}$ shows in-gap impurity states of both spin flavors. Since in this case the inversion symmetry is broken, the spin-degeneracy is, however, lifted. Upon close inspection, we see that the breaking of the spin-degeneracy due to broken inversion symmetry also holds for the bulk states in both cases.

For Cr interstitial vacancy complexes $V_{Cr}$+$Cr^{\delta 1}_{int}$ and $V_{Cr}$+$Cr^{\delta 2}_{int}$, we observe a similar behaviour, visible in Fig.~\ref{fig:theory}(c). In the case of $V_{Cr}$+$Cr^{\delta 2}_{int}$ we find nearly completely flat states, a remarkable result that indicates the highest localisation of the corresponding impurity states in real space. In contrast to the simple Cr vacancies, we see that the spin-degeneracy of the bulk states is strongly broken. Here, this results from charge transfer from the upper layer towards and into the vdW gap (following the Cr replacement). This strongly enhances the local inversion symmetry breaking. In the SM we show in detail how this effect is reduced by lowering the impurity density, i.e. by increasing the supercell size. The
$V_{Cr}$+$V_{Br}$ and $V_{Cr}$+$V_{2Br}$ leave similar footprints in the electronic structure, as seen in Fig.~\ref{fig:theory}(d). However, $V_{Cr}$+$V_{2Br}$ appears special as in this case a weakly dispersive impurity band crosses the Fermi level.

\begin{table}[htbp]
\begin{ruledtabular}
\begin{tabular}{ccccc}
Defect & $E^{bind}_{vdW}$ & $E^{bind}_{surface}$ & Point & Optical \\
Type & (meV) & (meV) & Group & activity \\
\hline
$V_{Cr}$ & 350 & 350 & $C_{2v}$ & Yes \\
$V_{S}$ & 279 & 278 & $C_{2v}$ & Yes \\
$V_{Br}$ & 161 & 165 & $C_{2v}$ & - \\
\hline
$2V^{bi}_{Cr}$ & 699 & - & $C_{2v}$ & Yes \\
$2V^{bi}_{S}$ & 557 & - & $C_{2v}$ & Yes \\
$2V^{bi}_{Br}$ & 325 & - & $C_{2v}$ & - \\
\hline
$V_{Cr}$+$V_{Br}$ & 300 & 306 & $C_{S}$ & Yes \\
$V_{Cr}$+$V_{2Br}$ & 282 & 293 & $C_{2v}$ & Yes \\
$V_{Cr}$+$V_{S}$ & 603 & 602 & $C_{S}$ & Yes \\
$V_{Cr}$+$V_{2S}$ & 710 & 707 & $C_{2v}$ & Yes \\
$V_{S}$+$V_{Br}$ & 402 & 407 & $C_{S}$ & Yes \\
\hline
\hline
Defect & $E^{bind}_{vdW}$ & $E^{bind}_{surface}$ & Point & Optical \\
Type & (meV) & (meV) & Group & activity \\
\hline
$V_{Cr}$+$Cr^{\delta 1}_{int}$ & 149 & 171 & $C_{S}$ & Yes \\
$V_{Cr}$+$Cr^{\delta 2}_{int}$ & 105 & 119 & $C_{S}$ & Yes \\
$V_{Cr}$+$Cr^{\delta 3}_{int}$ & 126 & 140 & $C_{1}$ & Yes \\
$V_{Cr}$+$Cr^{\delta 4}_{int}$ & 134 & 155 & $C_{S}$ & Yes \\
$V_{Cr}$+$Cr^{\delta 5}_{int}$ & 167 & 189 & $C_{S}$ & Yes \\
$V_{Cr}$+$Cr^{\delta 6}_{int}$ & 166 & 189 & $C_{1}$ & Yes \\
$V_{Cr}$+$Cr^{\delta 7}_{int}$ & 173 & 189 & $C_{1}$ & Yes \\
$V_{Cr}$+$Cr^{\delta 8}_{int}$ & 146 & 155 & $C_{1}$ & Yes \\
\end{tabular}
\end{ruledtabular}
\caption{\textbf{Parameters and properties of defects and defect complexes in 2L CrSBr.}
Binding energy densities, symmetry and potential optical activity based on occupied and unoccupied defect levels.
	}
\label{tab:structures}
\end{table}

We show the relaxed structures of all calculated defects and defect complexes in Fig.~\ref{fig:theory_structures} and their binding energy densities, electronic symmetry and potential optical activity in Tab.~\ref{tab:structures}. This emphasizes the fact that most of the defect complexes we have considered should show both occupied and unoccupied in-gap defect states. We can thus expect to find single-photon emitters among these defect structures, resulting from inter-defect optical excitations / relaxations. Together with their peculiar inherited spin states, this might allow for magnetically controlled single-photon emission. We finally note that the overall flatness of the in-gap defect states indicates a rather high degree of real-space localization of the corresponding defect states and thus possibly enhanced Coulomb interaction effects. We could therefore expect to find correlated / multi-reference behaviour among the defect complexes we have identified, suggesting that they need to be theoretically described beyond the mean-field level that we have applied here.

\section{Conclusion}

We have established a defect library in CrSBr by combining atomic resolution STEM, deep learning and theoretical ab-initio calculations. Due to its unusual crystal structure, CrSBr exhibits a large set of defects and defect complexes, including single and vertically stacked vacancies, interstitial-vacancy defect complexes and vacancy defect complexes. The results show that machine learning is effective in assisting the identification of defect types known to be present, particularly in scenarios where materials are beam sensitive and only very low doses can be used to obtain images. The use of deep learning enabled us to detect defect complexes that expand over more than one atom column, features that are otherwise challenging to resolve using conventional image analysis methods. However, by integrating and analyzing multiple images, more subtle features of the atomic arrangement could be uncovered. For CrSBr bilayers, this enhanced visualization showed that structures that appeared at first to be single vacancy defects were in fact more complex configurations. 

In particular, we detected $V_{Cr}$ defects associated with $Cr_{int}$ at specific locations predicted by DFT, and $V_{Cr}$ defects linked with $V_{Br}$ in a pattern explained by the geometric bond configuration within the structure. The interstitial defects are particularly compelling, as our results suggest that they create highly localized electronic states with optical transitions. Furthermore, detecting extended defect lines along the a-directions and a power-law distribution of their occurrence suggest correlated growth and provides an important framework to interpret defect-related transport measurements.

Our atomically precise measurement of these CrSBr defects and defect complexes creates an absolutely necessary foundation for further higher-level theoretical investigations of defect-induced electronic, magnetic, and optical properties. Based on our initial mean-field results indicating highly localized real-space defect states, we expect $V_{Cr}$+$Cr^{\delta 2}_{int}$ and $V_{Cr}$+$V_{2Br}$ to be especially interesting in potentially hosting (strongly) correlated properties.

Moreover, to the best of our knowledge, our work offers the first experimental atomic scale defect study that is representative for the structural FeOCl type with Pmmn space group symmetry. As such our results in CrSBr are a blueprint for defects such as interstitials and vacancy complexes that are similarly expected in more than 20 predicted compounds~\cite{Mounet.2018} such as the family of transition metal oxyhalides (MOX with M = Fe, Cr, V, Ti, Dy, Eu and X = Cl, Br, I)~\cite{Armand.1978,Mounet.2018} and transition metal chalcogenide halides (MChX with Ch = S, Se) similar to CrSBr, like CrSI, DySBr, ErSeI, ErSI and many more~\cite{Mounet.2018}.

Looking ahead, this work clarifies how the unique energetic landscapes of Cr atoms in CrSBr creates an environment that allows the formation of intriguing localized magneto-electronic defect states. The defects are expected to exhibit optical properties akin to quantum emitters in other layered materials~\cite{Montblanch.2023}. The presence of optically active defects is particularly exciting as CrSBr is known to maintain monolayer properties in bulk form~\cite{Klein.2023,Klein.2024b}, so the defects we discussed in bilayers of CrSBr can be straightforwardly extrapolated to the multilayer or bulk system. Indeed, bulk CrSBr has shown optically active defect emission for which the origin is still under debate~\cite{Klein.2022a}. Our detailed experimental and theoretical results suggest a plethora of optically active defects in CrSBr. The methodology we have developed to characterize defects in multilayer systems and the properties of defects, particularly related to Cr motivates future work to pursue atomic scale modification of CrSBr via focused electron beams~\cite{Klein.2022} for the deliberate generation of defect complexes relevant for quantum technologies.

%
%
\section{Acknowledgements}
J.K. and F.M.R. acknowledge funding through the NSF Trailblazer Award Number 2421694 and DOE-BES Award Number DE-SC0025387. The authors acknowledge the MIT SuperCloud and Lincoln Laboratory Supercomputing Center for providing HPC, database, and consultation resources that have contributed to the research results reported in this paper~\cite{reuther2018interactive}. Synthesis work at Columbia University was supported by the NSF through the Columbia University Materials Research Science and Engineering Center (MRSEC) on Precision-Assembled Quantum Materials DMR-2011738. Z.S. was supported by the ERC-CZ program (project LL2101) from Ministry of Education Youth and Sports (MEYS) and by the project Advanced Functional Nanorobots (reg. No. CZ.02.1.01/0.0/0.0/15\_003/0000444 financed by the EFRR). The Center for Visualizing Catalytic Processes (VISION) is funded by the Danish National Research Foundation (DNRF146). The computations were performed at the Dutch National Supercomputer Snellius under Project No. EINF-4184. This work was carried out in part through the use of MIT.nano’s facilities. We thank Matthias Florian for fruitful discussions. We acknowledge Referee 2 for the valuable suggestion that led us to extend the statistical analysis of vacancy defect lines.

\section{Author contributions}
J.K. conceptualized and supervised the project and together with M.W. designed the experiments. X.R., D.C., Z.S. and K.M. synthesized bulk crystals of CrSBr. J.K. prepared STEM samples. J.K. and A.P. collected STEM data. M.W. developed the machine learning work flow. S.G. and M.R. provided ab-initio DFT calculations. J.S., S.H. and F.M.R discussed the results. The manuscript was written by M.W. and J.K. with input from all co-authors.

%

\section{Appendix}

\subsection{Crystal growth}
Two separate sources of crystals were used and gave similar results. These crystals were fabricated using chemical vapor transport (CVT) as described in Refs.~\cite{Telford.2020} and~\cite{Klein.2022}.

\subsection{Sample fabrication}
Samples were fabricated by mechanical exfoliation of CrSBr onto SiO$_2$/Si substrates. Sample thickness was verified by atomic force microscopy, optical phase contrast and Raman spectroscopy. The samples were then transferred to S/TEM grids using dry polymer assisted transfer. No water was used for the transfer process to avoid additional surface damage due to hydrolysis~\cite{Torres.2023}. For additional information see the SI.

\subsection{Scanning transmission electron microscopy}
The STEM imaging was conducted on a probe-corrected Thermo Fisher Scientific Themis Z G3. The microscope was operated at a $\SI{200}{\kilo\electronvolt}$ acceleration energy, a beam current of $\SI{35}{\pico\ampere}$ and a semi-convergence angle of 19 mrad. This provided an aberration-corrected sub-Å probe size (typically 80 pm) allowing for atomic resolution. The images were all captured at 1024x1024 pixels with a pixel size of $\SI{7.8}{\pico\meter}$. We used a short pixel dwell time of $\SI{500}{\nano\second}$/px corresponding to a single image dose of $9 \cdot 10^5 \SI{}{\mathrm{e^-}\per\nano\meter\squared}$. Due to the structural sensitivity of bilayer CrSBr under the electron beam, we optimized image acquisition. We collected and averaged a series of 10 images using rigid registration~\cite{Savitzky.2018} with a total image dose of $9 \cdot 10^6 \SI{}{\mathrm{e^-}\per\nano\meter\squared}$. For each image series, it was necessary to adopt a specific acquisition strategy: We set a very low magnification to reduce dose, moved to a new and unexposed area and then immediately started a blind image acquisition at higher magnification. This approach ensured that the material was not exposed to the electron beam before data collection. As a result, most images are out of focus considering height variations across the sample but after post-selection a small fraction of images remained for our deep learning analysis.

\subsection{Image simulations}
HAADF-STEM image simulations were conducted with microscope parameters that matched the experimental conditions. The HAADF-STEM simulations used in the illustrations and in the datasets for training were constructed using abTEM~\cite{madsenAbTEMCodeTransmission2021}, a Python package for electron microscopy simulations. The simulations were accelerated using the plane wave reciprocal space interpolated scattering matrix (PRISM)~\cite{ophusFastImageSimulation2017} algorithm since STEM simulations are typically computationally expensive. During simulations of the training data several microscope aberrations were applied. 

For our comparison to the class averages of the defects we likewise used PRISM with detector collection angles of 78–200 mrad and a semi-convergence angle of 19 mrad and an interpolation factor of 1, but without any aberrations. We smoothed simulated images with a Gaussian full width at half maximum (FWHM) of $\SI{80}{\pico\meter}$ to match the experimental probe size and to account for thermal vibrations. Furthermore, the images were constructed by taking the average STEM-HAADF simulation between the two different sites within the unit cell where the defect could occur. The resulting line profiles were created by summing the intensity over the FWHM of the atoms, similarly done for both simulated and experimental data.

\subsection{Machine learning framework}
In this work we utilized a UNet++~\cite{zhouUNetNestedUNet2018} architecture with a depth of 5 to analyze HAADF-STEM images. We compared the UNet++ network to 7 other network architectures and found that for our use cases it consistently outperformed them including the UNet when tasked to find the position of atoms (see SI). The architecture is a derivative of the popular UNet~\cite{UNET} and provides additional pathways and an increased amount of blocks at several scales resulting in a more reliable information transfer. The training dataset was constructed using the Python package abTEM~\cite{madsenAbTEMCodeTransmission2021} and ASE~\cite{HjorthLarsen2017} to simulate images of 2L CrSBr with randomly introduced defects. The complete overview of the simulations and the training process is shown in the SI and the corresponding code is available on GitHub~\cite{Weile2024}.

\subsection{Ab-initio calculations}
For structure optimization and total energy calculations, we used the Vienna Ab Initio Simulation Package (VASP)~\cite{KRESSE199615,PhysRevB.54.11169} utilzing projector augmented plane waves (PAW)~\cite{paw,paw2} as the basis. 
To approximately address electronic correlations, we applied the rotationally invariant GGA+U approach with $U = 2.5$ eV and $J = 0.4$ eV (constrained random phase approximation values for single layer CrSBr in an effective dielectric environment of $\varepsilon_\text{env}\approx8$ ~\cite{rudenko_dielectric_2023}) acting within the Cr $d$-orbitals.
The PBE~\cite{PhysRevLett.77.3865} form of the GGA exchange-correlation potentials was used. 
For the simulation of defects in 2L structures, we utilized ($4\times3\times2$) supercells with a vacuum of 30 \AA, $2\times2\times1$ Monkhorst-Pack $k$-grids and a plane-wave cut-off energy of 500 eV.
The atomic positions are fully relaxed until all Hellmann-Feynman forces acting on each atom are reduced to (or below) 0.01 eV/\AA. 

%
%

\bibliographystyle{naturemag}
\bibliography{full}

\end{document}


\setcounter{figure}{0} 
\renewcommand{\thefigure}{S\arabic{figure}}

\renewcommand{\thetable}{S\arabic{table}}

\fontsize{11pt}{13pt}\selectfont

\makeatletter
\renewcommand\@make@capt@title[2]{%
    \@ifx@empty\float@link{\@firstofone}{\expandafter\href\expandafter{\float@link}}%
    \fontsize{11pt}{13pt}\selectfont\textbf{#1}\@caption@fignum@sep#2
}
\renewcommand\figurename{Figure}
\makeatother

\title{Supplementary Material: \\ Defect complexes in CrSBr revealed through electron microscopy and deep learning}
%
%
\author{Mads~Weile}\email{weile.a.mads@gmail.com}
\affiliation{Department of Materials Science and Engineering, Massachusetts Institute of Technology, Cambridge, Massachusetts 02139, USA}
\affiliation{Center for Visualizing Catalytic Processes (VISION), Department of Physics, Technical University of Denmark, 2800 Kongens Lyngby, Denmark}
\author{Sergii~Grytsiuk}
\affiliation{Institute for Molecules and Materials, Radboud University, Heijendaalseweg 135, 6525AJ Nijmegen, The Netherlands}
\author{Aubrey~Penn}
\affiliation{MIT.nano, Massachusetts Institute of Technology, Cambridge, Massachusetts 02139, USA}
\author{Daniel~G.~Chica}
\affiliation{Department of Chemistry, Columbia University, New York 10027, United States}
\author{Xavier~Roy}
\affiliation{Department of Chemistry, Columbia University, New York 10027, United States}
\author{Kseniia~Mosina}
\affiliation{Department of Inorganic Chemistry, University of Chemistry and Technology Prague, Technická 5, 166 28 Prague 6, Czech Republic}
\author{Zdenek~Sofer}
\affiliation{Department of Inorganic Chemistry, University of Chemistry and Technology Prague, Technická 5, 166 28 Prague 6, Czech Republic}
\author{Jakob~Schiøtz}
\affiliation{Department of Physics, Technical University of Denmark, DK-2800 Kgs., Lyngby, Denmark}
\author{Stig Helveg}
\affiliation{Center for Visualizing Catalytic Processes (VISION), Department of Physics, Technical University of Denmark, 2800 Kongens Lyngby, Denmark}
\author{Malte~Rösner}
\affiliation{Institute for Molecules and Materials, Radboud University, Heijendaalseweg 135, 6525AJ Nijmegen, The Netherlands}
\author{Frances~M.~Ross}\email{fmross@mit.edu}
\affiliation{Department of Materials Science and Engineering, Massachusetts Institute of Technology, Cambridge, Massachusetts 02139, USA}
\author{Julian~Klein}\email{jpklein@mit.edu}
\affiliation{Department of Materials Science and Engineering, Massachusetts Institute of Technology, Cambridge, Massachusetts 02139, USA}
%
%

%
\maketitle
%
%

\tableofcontents

\newpage

%
\begin{figure*}[ht]
\scalebox{\figurescale}{\includegraphics[width=1\linewidth]{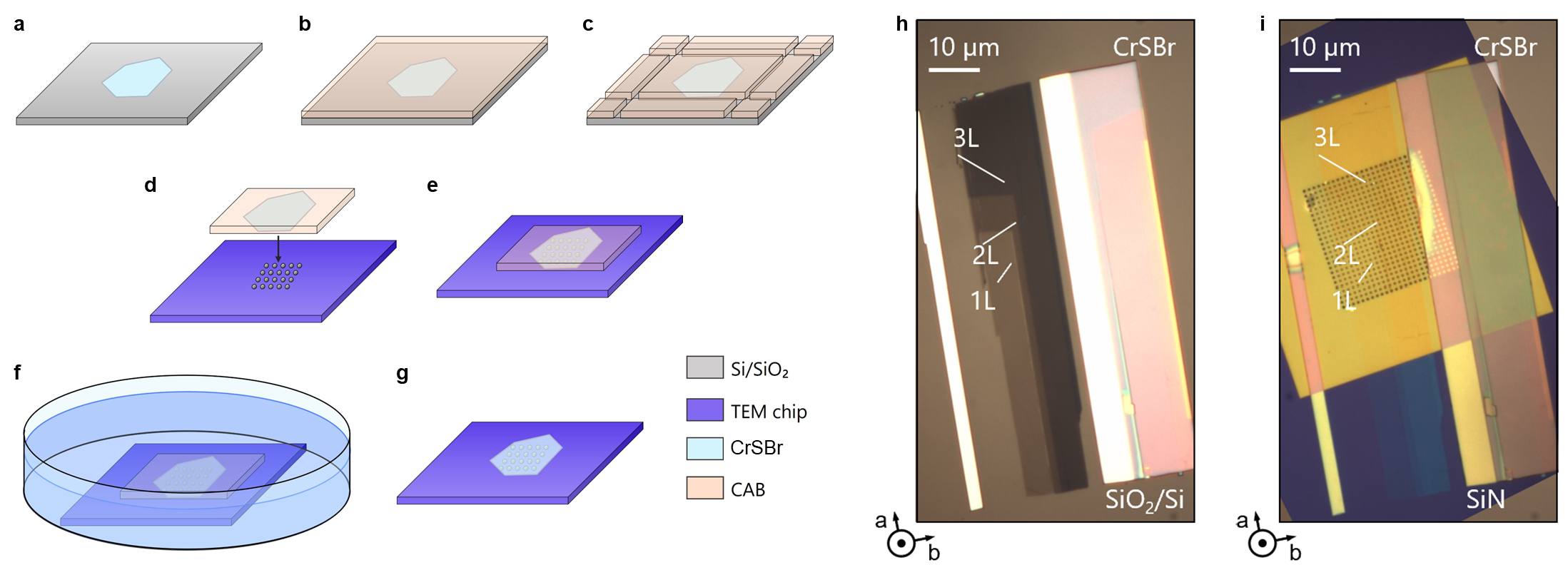}}
\caption{\label{fig:sample_preparation}
\textbf{Transmission electron microscopy sample fabrication process.} 
\textbf{a}, exfoliated flakes of CrSBr on a SiO$_2$/SM chip.
\textbf{b}, spin-coated chip with the polymer cellulose acetate butyrate. 
\textbf{c}, stamp cut-out.
\textbf{d}, stamp lift-off and transfer.
\textbf{e}, stamp baking.
\textbf{f}, submersion in acetone bath followed by an IPA bath.
\textbf{g}, resulting sample after Critical Point Drying.
\textbf{h}, optical micrograph image of exfoliated CrSBr flakes on SiO$_2$/Si chip.
\textbf{i}, optical micrograph image of transferred CrSBr flakes on TEM chip.
}
\end{figure*}

\section{1. Sample Preparation for Scanning Transmission Electron Microscopy}

\textbf{Fig.~\ref{fig:sample_preparation}a-i} schematically depicts the S/TEM sample fabrication process step-by-step. We begin by exfoliating CrSBr flakes onto a SiO$_2$/Si chip (\textbf{Fig.~\ref{fig:sample_preparation}a}). After this, we spin-coated the chip with a thin layer of the polymer cellulose acetate butyrate (CAB) (\textbf{Fig.~\ref{fig:sample_preparation}b}). CAB is used due to its low glass transition temperature of $\approx 130^\circ$C that allows for baking while avoiding flake damage in the process~\cite{schneiderWedgingTransferNanostructures2010}. Specifically, a solution of CAB in ethyl acetate 24g/100ml is used. The CAB is drop-casted onto the SiO$_2$/Si chip. The chip is spin-coated at 1500 rpm for 60s with an acceleration of 500 rpm/s which results in a polymer thickness of approximately $\SI{15}{\micro\meter}$. After spin coating, the polymer is baked at 80$^\circ$C for 6 minutes with a clean petri dish serving as a lid to ensure an even bake preventing the top layer of the CAB from drying out. Next, using a scalpel mm sized area is cut around desired flakes under an optical microscope (\textbf{Fig.~\ref{fig:sample_preparation}c}). This cut out stamp is mechanically moved with a tweezer and placed onto a holey grid on a S/TEM chip under an optical microscope (\textbf{Fig.~\ref{fig:sample_preparation}d, e}). After the stamp is transferred onto the S/TEM chip, the entire chip is baked at 80 $^\circ$C for 1 minute. This is above the CAB glass transition temperature causing the CAB to conform to the surface improving the adhesion of flakes. The last step of the process is to remove the CAB using solvents (\textbf{Fig.~\ref{fig:sample_preparation}f}). Here the sample is first submerged in acetone for 15 minutes after which it is submerged into IPA to remove the leftover acetone. In a last step, a critical point dryer (CPD) is used where the sample submerged in IPA is sealed in the CPD chamber. The chamber is then flooded by liquid CO$_2$ and the transition from supercritical fluid to gas is utilized avoiding surface tension. The overall process results in high quality samples with minimized sample damage on the atomic scale and reduced surface contamination. This is particularly important for CrSBr that is susceptible to damage when in contact with water~\cite{Torres.2023}.
An optical microscope image of an exfoliated CrSBr flake on a SiO$_2$/Si chip is shown in \textbf{Fig.~\ref{fig:sample_preparation}h} and the same flake after the described polymer-assisted transfer is shown in \textbf{Fig.~\ref{fig:sample_preparation}i}.

\section{2. Microscopic Structural Stability of Mono- and Few-Layer CrSBr}

\begin{figure*}[ht]
\scalebox{\figurescale}{\includegraphics[width=1\linewidth]{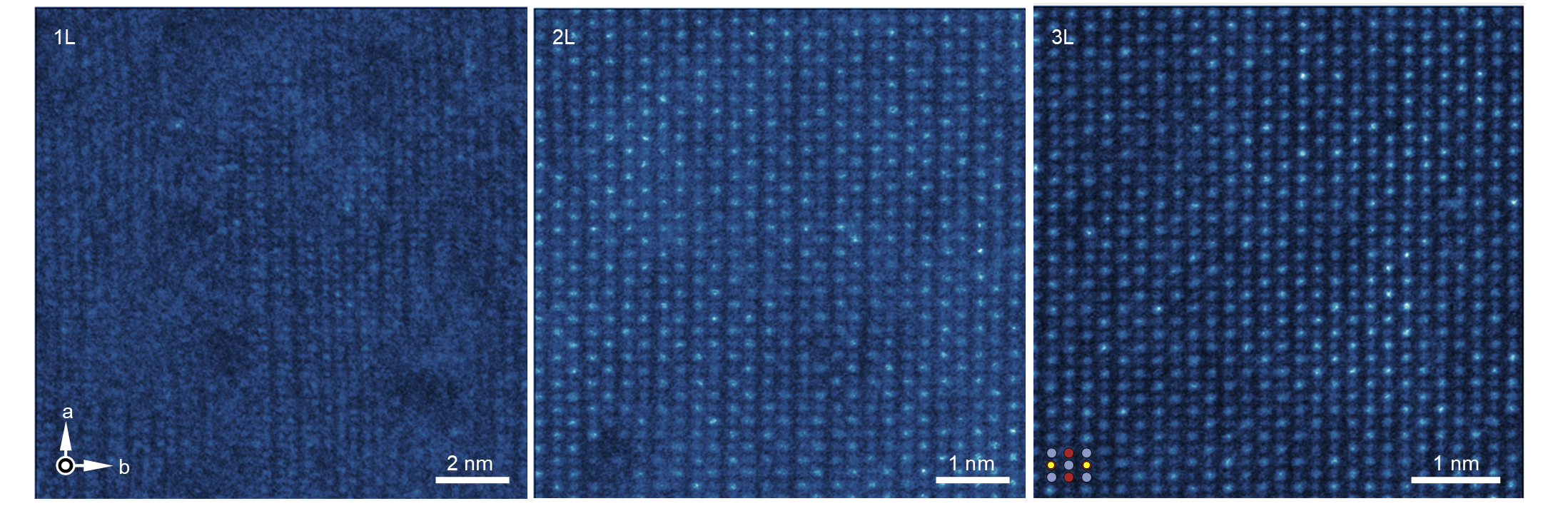}}
\caption{\label{fig:stabilty_CrSBr}
\textbf{Structural stability of CrSBr.} 
HAADF-STEM images of 1L, 2L and 3L CrSBr. The 1L CrSBr shows significantly more structural damage as compared to 2L or thicker material. The beam energy is $\SI{200}{\kilo\electronvolt}$ and the beam current is $\SI{35}{\ampere}$. The total dose for each image is $9 \cdot 10^6 \SI{}{\mathrm{e^-}\per\nano\meter\squared}$.
}
\end{figure*}

The microscopic stability of CrSBr is essential for conducting defect studies in STEM. Typical HAADF-STEM images of 1L, 2L and 3L CrSBr are shown in \textbf{Fig.~\ref{fig:stabilty_CrSBr}}. While the bilayer (2L) and thicker material show good structural stability, the 1L exhibits significant sample damage and structural distortions with large, nm-sized holes and disintegration of parts of the sample~\cite{Torres.2023}. We therefore focus our study on 2L CrSBr.

%
%

\section{3. Formation of Cr Interstitial Defects Under Electron Beam Irradiation}

\begin{figure*}[ht]
\scalebox{\figurescale}{\includegraphics[width=0.9\linewidth]{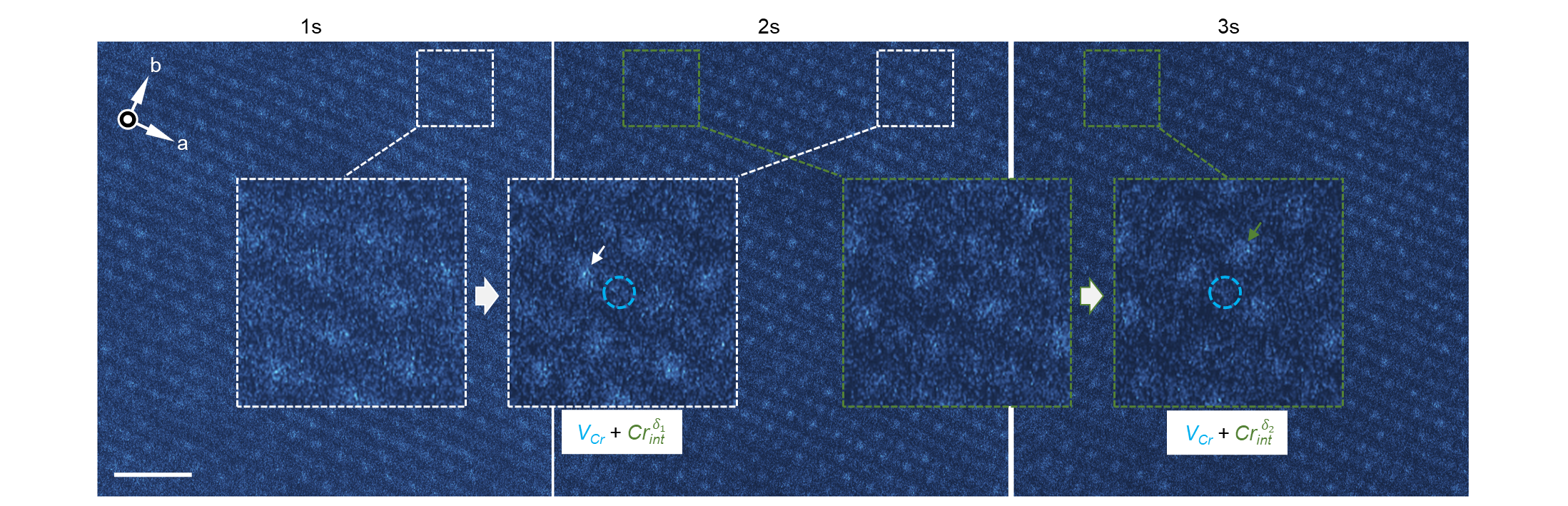}}
\caption{\label{fig:time_dependence}
\textbf{Formation of Cr interstitial defects under electron beam irradiation in 2L CrSBr.} 
HAADF-STEM images of 2L CrSBr showing the formation of $V_{Cr}$+$Cr^{\delta 1}_{int}$ (white) and $V_{Cr}$+$Cr^{\delta 2}_{int}$ (green). Each image is the average of two images with an integration time of $\SI{0.5}{\second}$. The beam energy is $\SI{200}{\kilo\electronvolt}$ and the beam current is $\SI{35}{\ampere}$. The total dose for one averaged $\SI{0.5}{\second}$ image is $1.8 \cdot 10^6 \SI{}{\mathrm{e^-}\per\nano\meter\squared}$. Scale bar is $\SI{1}{\nano\meter}$.
}
\end{figure*}

Exposure of bulk CrSBr with high energy electrons has recently shown the migration of Cr atoms into interstitial sites between the van der Waals gaps resulting in the formation of a new crystal phase~\cite{Klein.2022}. The movement appears highly selective for the Cr atoms and occurs at a high electron beam energy of $\SI{200}{\kilo\electronvolt}$. In our experiments on 2L CrSBr, we collected an image series of 10 consecutive images to study the structural changes during exposure where each image has an exposure time of $\SI{0.5}{\second}$. \textbf{Fig.~\ref{fig:time_dependence}} shows three HAADF-STEM images, each the sum of 2 consecutive frames resulting in a total image time of $\SI{1}{\second}$. During imaging, the creation of Cr vacancies and simultaneous movement of the Cr atom to a proximal atom columns is observed throughout all data. We highlight the formation of two characteristic interstitial vacancy complexes that form between consecutive frames. While we point out the formation of $V_{Cr}$+$Cr^{\delta 2}_{int}$ and $V_{Cr}$+$Cr^{\delta 1}_{int}$, other Cr movements into S/Br and Cr atom columns occurs.

%
%


\section{4. Machine Learning work flow}

To study the defects in 2L CrSBr we develop a machine learning work flow. This work flow can be directly applied for any other desired 2D material, defect types and layer thicknesses. Generally, The performance of the resulting models depends largely on the quality and representatives of the data used during training~\cite{deeplearningaiChatAndrewMLOps2021, ghoshEnsembleLearningiterativeTraining2021a}. In our work flow we adopt an approach based on simulated data inspired by \cite{madsenDeepLearningApproach2018,ziatdinovDeepLearningAtomically2017}. In the following, each step of the work flow is described in detail. The code to replicate and use this exact work flow is available on GitHub \url{https://github.com/MadsAW/2d-defect-detection} with accompanying Python notebooks demonstrating the different work flow steps.

%
\begin{figure*}[ht]
\scalebox{\figurescale}{\includegraphics[width=0.6\linewidth]{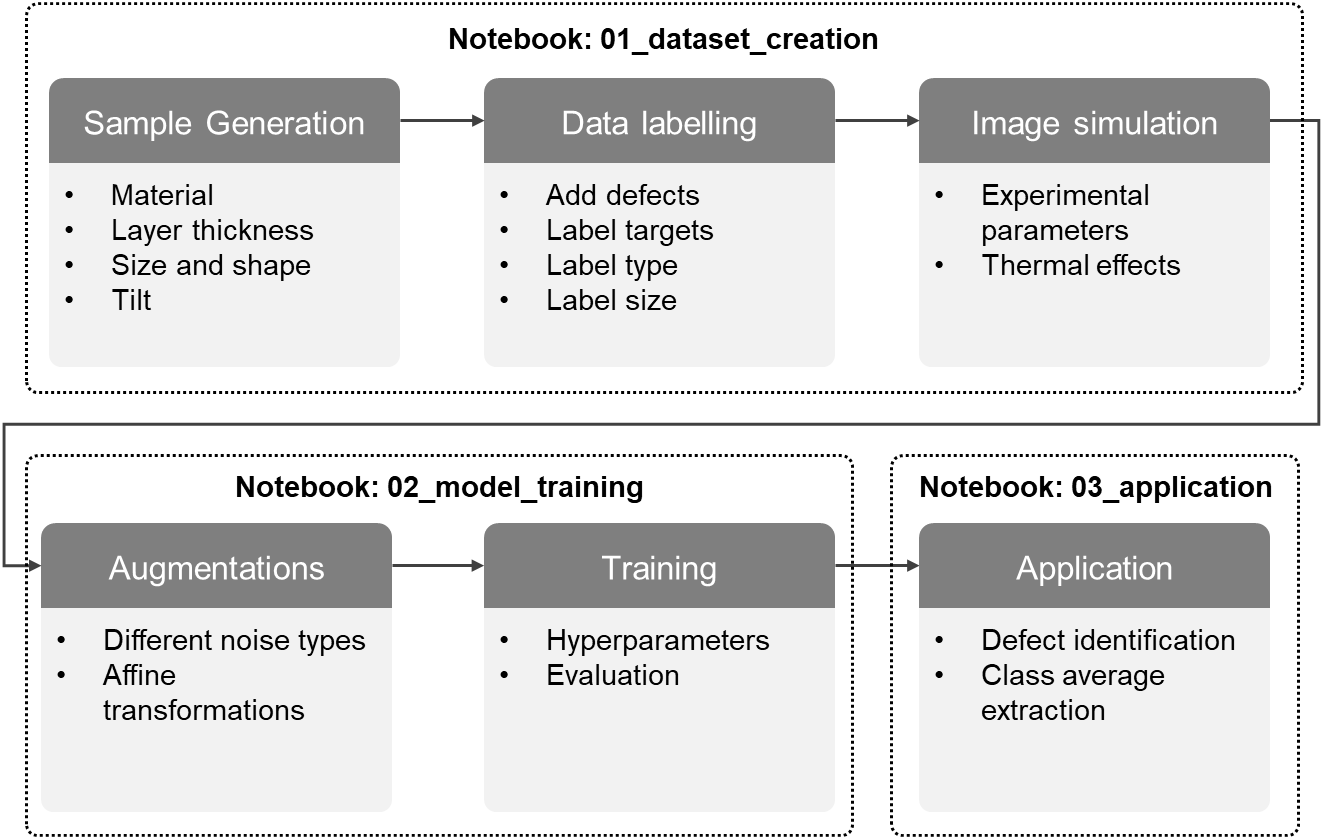}}
\caption{\label{fig:overivew}
\textbf{Machine learning work flow.} The above work flow demonstrates the different components of the work flow and their elements. The first three steps are used to create the dataset whereas the next two steps are used when training a new model. The last step and notebook represents the analysis conducted on experimental data using the trained networks.
}
\end{figure*}
%

The first step in the work flow is to construct digital samples of the desired material, these are referred to as atomic models. Secondly, specified defects are introduced in the atomic models and their locations are added to a label representing the ground-truth location of defects. Third, HAADF-STEM images are simulated with appropriate experimental parameters. Fourth, the simulated images are augmented to increase the dataset size while including different noise sources and affine transformations. Fifth, the resulting datasets of labels and simulated images are used to train deep convolutional neural networks (DCNNs) which are evaluated on experimental data. Lastly, the networks are applied to experimental images to find defects and extract class-averages. Below each of the different steps are described in further detail.

\subsection{4.1. Sample Generation} 

The first step is to create an atomic model of the material of interest. This model is a structural model of atomic species and corresponding positions. The sample is used for two purposes: Simulating an electron microscopy image and for creating a label. To generate atomic models of our samples we use the atomic simulation environment (ASE) python package~\cite{larsenAtomicSimulationEnvironment2017}.

\begin{figure*}[ht]
\scalebox{\figurescale}{\includegraphics[width=0.8\linewidth]{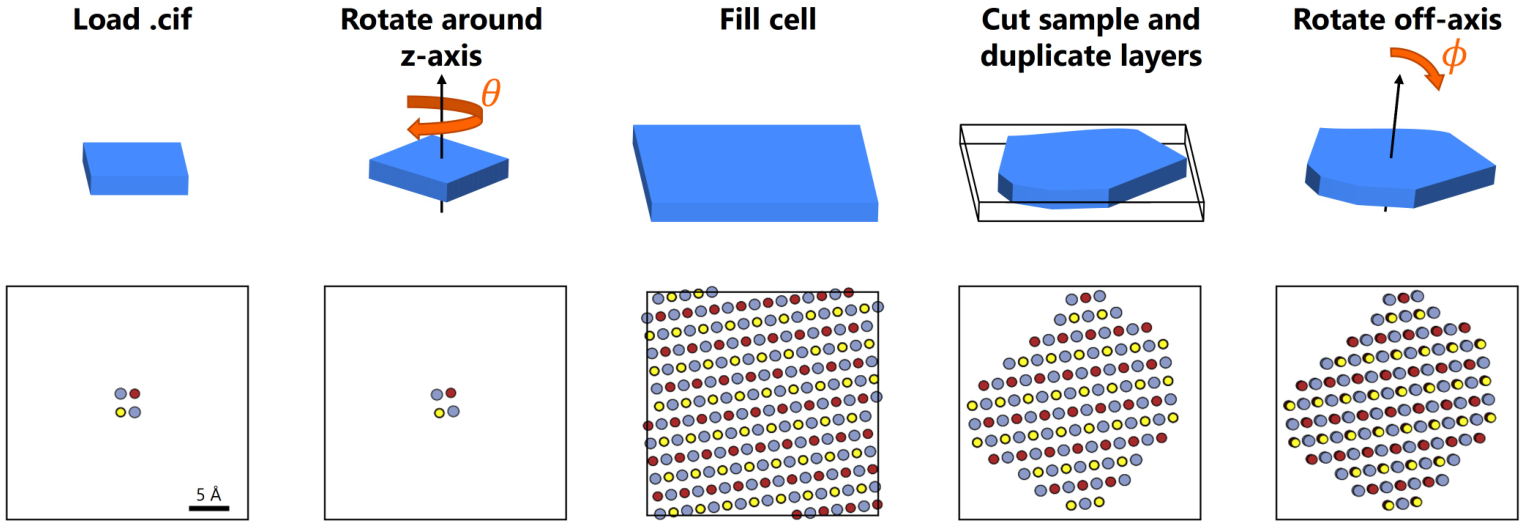}}
\caption{\label{fig:sample_generation}
\textbf{Sample generation process.} 
The upper row schematically depicts the sample generation process, whereas the lower row shows the atomic model of a 2L CrSBr sample in a top-down view of the x-y-plane. First, the .cif file of the relevant compound is loaded. The loaded unit cell is then randomly rotated around the z-axis. The unit cell is replicated in-plane to fill the extent of the cell, and multiplied out-of-plane by the desired layer number. The sample is randomly cut along several directions to create a flake with random edges. Finally, the sample is randomly rotated off-axis by up to a few degrees.
}
\end{figure*}

The atomic model generation process is depicted in \textbf{Fig.~\ref{fig:sample_generation}}. In our work flow we generate atomic models of flakes of van der Waals (vdW) materials with varying sizes (x-y-plane) and layer numbers (z-direction). First, an input .cif file is used that represents the unit cell of the material. The unit cell is randomly rotated in-plane, $\theta\in [0,360]$, and replicated to achieve a desired cell size, thereby creating a square sample. The sample is then replicated along the z-direction to generate a desired layer number. Next, the square sample is randomly sliced along different directions $\{ \langle i, j, 0 \rangle \mid i, j \in \{-1, 0, 1\}, (i,j) \neq (0,0) \}$, resulting in the generation of convex polygon flakes. Lastly, to account for the possibility that not every image will be "on-zone", the samples are rotated randomly off-axis by up to a few degrees, $\phi\in [0,1.6]$. Generating samples allows for a lot of flexibility in the size of the flake, layer number, and rotation angle. These are sample generation parameters that are unique to a given dataset.

%
%

The resulting sample is represented by the atomic model. This model can be directly changed within ASE for example with the introduction of defects. The models can then be passed directly to abTEM~\cite{madsenAbTEMCodeTransmission2021} to simulate an ideal image of the sample. Note, by using flakes for the atomic model we automatically introduce different edges and interfaces into the dataset which helps to prevent overfitting to perfect lattices. Likewise, we can also slightly shift each atomic column randomly further distorting the lattice. It is important to note that the vacuum around the sample is needed to perform accurate simulations. This is because the Fourier transform used in the multislice and PRISM simulation assumes that the periodicity wraps around the entire object.

\subsection{4.2. Data Labelling} 

\begin{figure*}[ht]
\scalebox{\figurescale}{\includegraphics[width=0.6\linewidth]{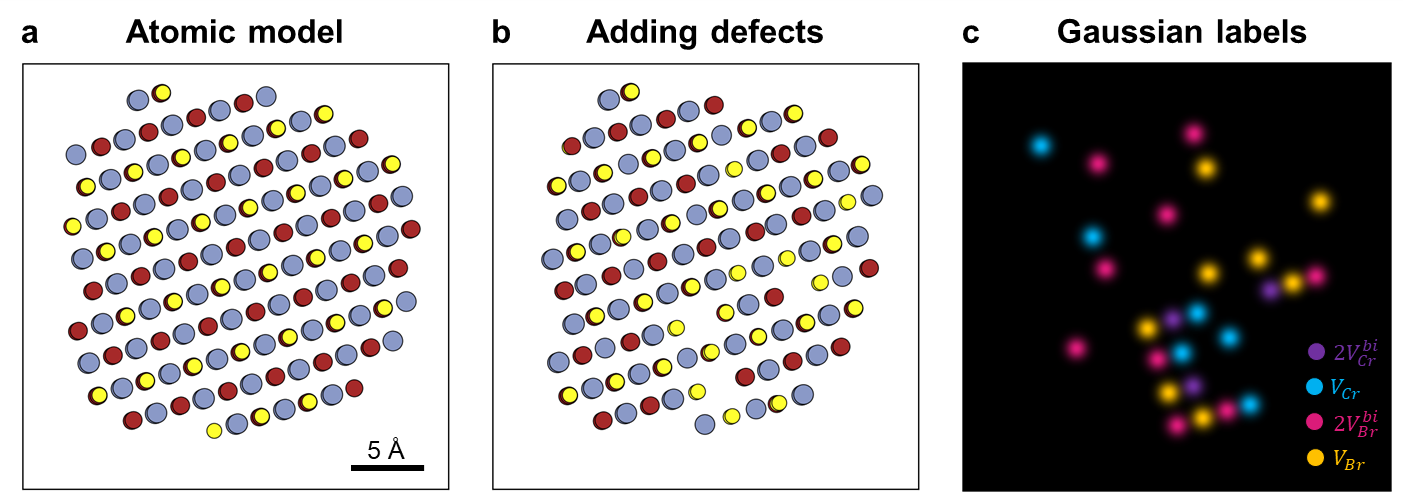}}
\caption{\label{fig:labeling}
\textbf{Introduction of defects and representation as a segmentation map for a flake of CrSBr} 
\textbf{a}, Top-down view of the atomic model for a flake of CrSBr. 
\textbf{b}, The same flake after the different defects have been introduced. 
\textbf{c}, The corresponding Gaussian label, here different colors respond to different defect types. 
}
\end{figure*}

In this step the different defects of interest are \textit{column-wise} introduced into the atomic model. For this work we used a 15\% probability for each of the different defects, i.e. the chance to encounter a $V_{Cr}$ in a Cr-column is 15\%. Here the locations of each of the defects are recorded and are translated into a segmentation image by introducing a labelling function that is fitted to each point and summed up. Each type of defect becomes a different channel in the label. An example of introducing defect and generating a Gaussian label, $\sigma = 0.5 \ \text{\AA}$, is depicted in \textbf{Fig.~\ref{fig:labeling}}. In our case we have chosen to use Gaussian label functions enabling sub-pixel refinement using their center-of-mass and we also construct a background label (channel) ensuring the normalization of the label.

%
%

\subsection{4.3. Image Simulation} 

Having created an atomic model of the sample an ideal image can be simulated. To this end we utilize abTEM~\cite{madsenAbTEMCodeTransmission2021}, which is a Python package specialized in simulating electron microscopy images under a variety of different conditions, both for TEM and STEM. In our case we simulate STEM images and make use of the PRISM algorithm that enables faster simulations~\cite{ophusFastImageSimulation2017}, here using an interpolation factor of 4. 

\begin{figure*}[ht]
\scalebox{\figurescale}{\includegraphics[width=0.8\linewidth]{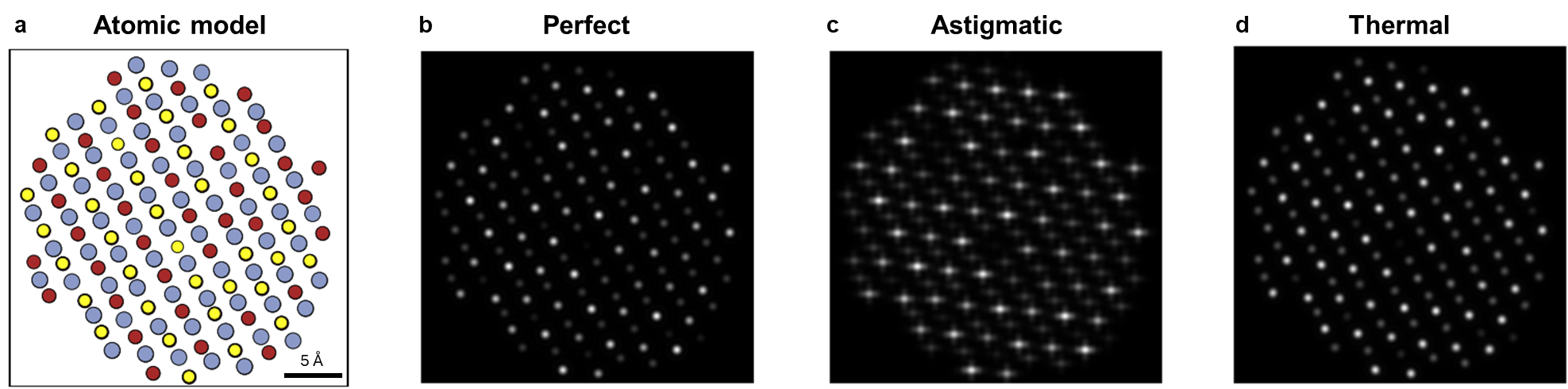}}
\caption{\label{fig:simulate_ideal_image_STEM}
\textbf{HAADF-STEM simulations of a flake of CrSBr.} 
\textbf{a}, Top-down view of a flake of CrSBr used for the STEM simulations. The simulations use an energy of 200 keV. 
\textbf{b}, HAADF-STEM simulation with no aberrations. 
\textbf{c}, HAADF-STEM simulation with astigmatism, here the atoms appear as streaks.
\textbf{d}, HAADF-STEM simulation with thermal effects corresponding to room temperature, here the atoms appear smeared. The effect of temperature can be simulated by blurring the resulting image.
}
\end{figure*}

 The image formation process in an electron microscope is not perfect and will contain a varying degree of aberrations due to imperfect lenses, settings, and the environment. To model these different conditions each image is simulated using random aberration coefficients within a range expected of experiments. The distributions used in this manuscript are depicted in \textbf{Tab.~\ref{tab:parameters_table}}. Some simulations depicting common aberrations are depicted in \textbf{Fig.~\ref{fig:simulate_ideal_image_STEM}}.

%

\subsection{4.4. Noise and Augmentations}

To utilize the simulations to their full extent we perform data augmentation in which the final dataset contains several augmented version of each simulated image. These transformations serve two purposes, first, to increase the dataset in a computationally cheap fashion and second, to model the experimental data. 

In this work flow, we will use two different types of augmentations. First, augmentations that model different noise sources present in experimental images. Second, affine transformations like cropping, flipping, and rotating. Affine transformations are commonly used in data augmentation of visual data and must be applied on both the image and label. Lastly, the input image is normalized before being used in the neural network. For our networks we apply 40 augmentation to each image resulting a final training set size of 24000 HAADF-STEM images and corresponding labels. Below we describe the different noise sources we model and transformations in further detail.

\subsubsection{4.4.1. Surface Contamination} 

In STEM experiments surface contamination is almost always present due to sample preparation or microscope conditions. Surface contamination can have both a static and a dynamic component. Static contamination is commonly present as a distribution of hydro-carbons on the sample surface~\cite{Lin.2011,Dyck.2021}. The most prominent dynamic surface contamination is electron-beam-induced deposition (EBID)~\cite{Dyck.2017}. EBID is observed as a background that builds under electron beam illumination as hydro-carbons on the sample are attracted from the surrounding and polymerized under the beam. This manifests in bright patches in HAADF-STEM. Attempts at minimizing contamination should be done during sample preparation. Even then, the contamination can also be addressed by modeling the EBID deposition in our training data.

There are several different approaches to including contamination in the synthetic data. The most common approach is to include a linear or parabolic background~\cite{linTEMImageNetTrainingLibrary2021}. Another approach is to create a library of background by high-pass filtering experimental images~\cite{leeDeepLearningEnabled2020,leeSTEMImageAnalysis}. Instead of these two techniques, we instead model contamination using Perlin noise~\cite{eeveePerlinNoise2016}.

%
\begin{figure*}[ht]
\scalebox{\figurescale}{\includegraphics[width=0.7\linewidth]{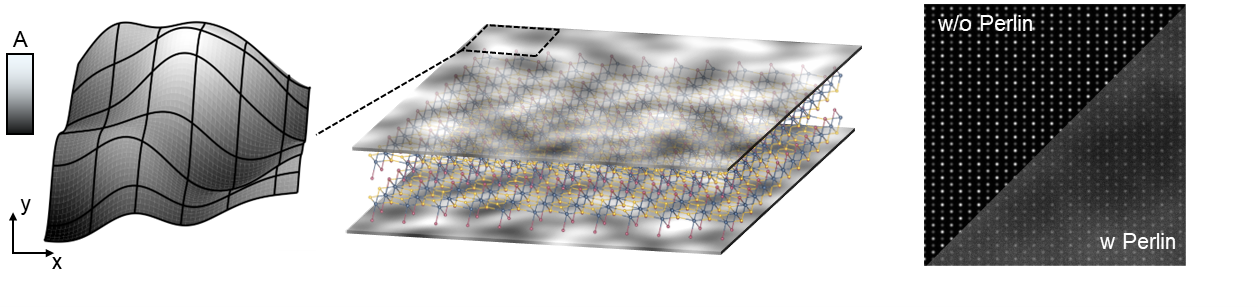}}
\caption{\label{fig:contamination}
\textbf{Modelling static and dynamic EBID contamination with Perlin Noise.} 
Hydrocarbon contamination can be modelled as a 3D landscape deposited on top of the sample. To simulate the effect of EBID contamination, a background of Perlin noise can be added to the simulated images.
}
\end{figure*}
%

Our approach assumes that the contamination can be described as a three-dimensional (3D) landscape of hydro-carbons deposited on top of our sample with a homogeneous scattering cross-section. This implies that the intensity of the formed image in HAADF-STEM can be calculated as the sum of the contamination landscape and the original HAADF-STEM image: $I_{\text{after}}(x,y) = I(x,y) + N_{\text{Perlin}}(x,y)$. We can simulate such a contamination landscape using Perlin noise. Perlin noise is a fractal noise type that is used often in computer graphics to generate landscapes and textures. This approach is illustrated in \textbf{Fig.~\ref{fig:contamination}}.

\subsubsection{4.4.2. Instrument Noise Sources} 

When acquiring experimental images there exist a variety of different noise sources. In our work flow we implement the following noise sources that can affect an input image, $I(x,y)$:

\begin{itemize}
    \item \textbf{Shot noise:} The most common noise present in all electron microscopy data is shot noise. Here the image formed on the detector follows a Poisson distribution since only a discrete amount of electrons can be detected. 
    $I_{\text{after}}(x,y)\sim N_{\text{Poisson}}(I(x,y))$
    \item \textbf{Image blurring:} A common effect in images is blurring either due to thermal atomic movement or because of drift in the microscope that when corrected slightly blurs the image. To take this into account the image can be slightly blurred. $I_{\text{after}}(x,y) = \frac{1}{2\pi\sigma^2} \exp\Big(-\frac{x^2+y^2}{2\sigma^2}\Big) I(x,y)$
\end{itemize}

Each noise source presents new parameters that can be altered. By tuning these, the synthetic data become more similar to experimental data.

\subsubsection{4.4.3. Affine Transformations} 

In our work flow we use the following affine transformations:

\begin{itemize}
    \item \textbf{Flip}: Randomly flip, mirror, the image along the x- and/or y-axis.
    \item \textbf{Rotate}: Randomly rotate the image by 90 degrees.
    \item \textbf{Resize and crop}: Resize the image to a given size and crop out an area of a specific size.
\end{itemize}

Here by simply flipping and rotating a single image, we would obtain 8 similar images. Afterward, we can resize the image to achieve different sampling resolutions and crop out an area of a given size. The affine transformations are applied to both the simulated images and the corresponding labels. Furthermore, the resizing augmentation helps generalize the network to a wider range of magnifications.

\subsection{4.5. Hyperparameters Overview}

To create realistic data for training the defect finding DCNNs, we used a range of microscope parameters summarized in \textbf{Tab.~\ref{tab:parameters_table}}. Our choice of parameters is based on the experimental conditions of the microscope for collecting HAADF-STEM images. Moreover, we vary certain parameters to account for variations in the experimental conditions such as aberrations or electron dose.

\begin{table}[h]
\centering
\begin{tabular}{|llll|}
\hline
\multicolumn{1}{|l|}{ \textbf{Parameter}} & \multicolumn{1}{l|}{{\textbf{Lower bound}}} & \multicolumn{1}{l|}{ \textbf{Upper bound}} & { \textbf{Distribution}} \\ \hline
\multicolumn{4}{|l|}{\textbf{Atomic model parameters}} \\ \hline
\multicolumn{1}{|l|}{Size} & \multicolumn{3}{l|}{30$\times$30 \AA$^2$} \\ \hline
\multicolumn{1}{|l|}{Layer number} & \multicolumn{3}{l|}{2} \\ \hline
\multicolumn{1}{|l|}{Col. displacement} & \multicolumn{3}{l|}{Norm($\mu=$0 \AA, $\sigma=$0.1 \AA)} \\ \hline
\multicolumn{4}{|l|}{\textbf{Simulation parameters}} \\ \hline
\multicolumn{1}{|l|}{Energy} & \multicolumn{3}{l|}{200 keV} \\ \hline
\multicolumn{1}{|l|}{Scanning sampling} & \multicolumn{3}{l|}{0.0625 \AA} \\ \hline
\multicolumn{1}{|l|}{Detector collection angles} & \multicolumn{3}{l|}{78 mrad to 200 mrad} \\ \hline
\multicolumn{1}{|l|}{Semiangle cutoff} & \multicolumn{1}{l|}{12 mrad} & \multicolumn{1}{l|}{27 mrad} & Uniform \\ \hline
\multicolumn{1}{|l|}{Defocus, $\Delta f$} & \multicolumn{1}{l|}{-3 nm} & \multicolumn{1}{l|}{3 nm} & Uniform \\ \hline
\multicolumn{1}{|l|}{Spherical, Cs} & \multicolumn{1}{l|}{0} & \multicolumn{1}{l|}{3 µm} & Uniform \\ \hline
\multicolumn{1}{|l|}{Astigmatism, C12} & \multicolumn{1}{l|}{0} & \multicolumn{1}{l|}{4 nm} & Uniform \\ \hline
\multicolumn{1}{|l|}{Coma, B2} & \multicolumn{1}{l|}{0} & \multicolumn{1}{l|}{50 nm} & Uniform \\ \hline
\multicolumn{4}{|l|}{\textbf{Experimental transformations}} \\ \hline
\multicolumn{1}{|l|}{Dose} & \multicolumn{1}{l|}{$1\cdot10^4 \ \mathrm{\frac{e^-}{\text{\AA}^2}}$ } & \multicolumn{1}{l|}{$5\cdot10^5 \ \mathrm{\frac{e^-}{\text{\AA}^2}}$} & Exponential \\ \hline
\multicolumn{1}{|l|}{Blur} & \multicolumn{1}{l|}{2} & \multicolumn{1}{l|}{4} & Uniform \\ \hline
\multicolumn{1}{|l|}{Contamination strength} & \multicolumn{1}{l|}{0} & \multicolumn{1}{l|}{0.4} & Uniform \\ \hline
\multicolumn{1}{|l|}{Background strength} & \multicolumn{1}{l|}{0.0005} & \multicolumn{1}{l|}{0.005} & Uniform \\ \hline
\multicolumn{4}{|l|}{\textbf{Affine transformations}} \\ \hline
\multicolumn{1}{|l|}{Resize scale} & \multicolumn{1}{l|}{0.8} & \multicolumn{1}{l|}{1.2} & Uniform \\ \hline
\multicolumn{1}{|l|}{Resize ratio, elongation} & \multicolumn{1}{l|}{0.9} & \multicolumn{1}{l|}{1.2} & Uniform \\ \hline
\end{tabular}
\caption{\textbf{Parameters used for generating data for the CrSBr defect detection ensemble}. The atom spotting network is trained on similar but slightly broader simulation parameter ranges in order to generalize better. Likewise these networks are also trained on a range of layer numbers.}
\label{tab:parameters_table}
\end{table}

\subsection{4.6. Training the Model}\label{sec:training} 

Using the generated training set we can train neural networks. To define and train our neural networks we use PyTorch. The networks are trained on electron microscopy images resized to $256 \times 256$ pixels and a label consisting of five channels with Gaussian labels: $V_{Br}$, $V_{Cr}$, $2V^{bi}_{Br}$ and $2V^{bi}_{Cr}$ and a background. 

%
\begin{figure*}[ht]
\scalebox{\figurescale}{\includegraphics[width=\linewidth]{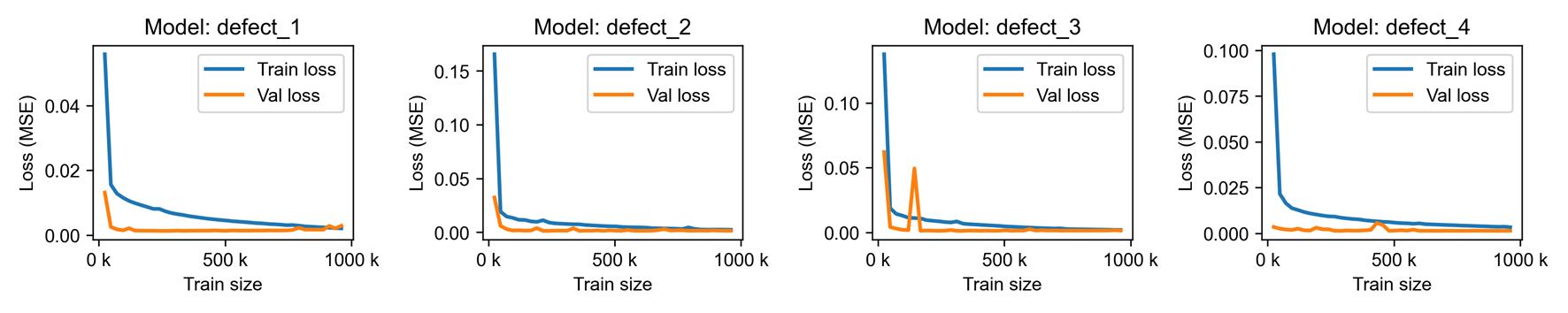}}
\caption{\label{fig:training}
\textbf{Learning curve for an ensemble of networks trained to detect defects in 2L CrSBr.} 
For each of the different networks in the ensemble the curve depicts the MSE loss evaluated on both the training set and validation set. As seen the curves converge. The validation set generally shows lower loss due to regularization which is turned off during evaluation of the validation dataset. The parameters used for training the network are available in \textbf{Tab.~\ref{tab:parameters_table}}.
}
\end{figure*}
%

For training the neural networks, we use an Adam optimizer with a learning rate of 0.0005, mean squared loss (MSE), as our loss function, a UNET++ network architecture with a depth of 5 and apply transfer learning where pre-trained weights for the encoder are used to initialize our network. The weights were pre-trained on ImageNet~\cite{imagenet}. We note that we use a shallow depth of the network to mitigate overfitting, which is one of the primary challenges when training networks on synthetic data that are later applied to experimental data. We refer to this as a deep neural network, even though the depth is only 5, to follow the conventions used in the field. The networks are trained on Nvidia Volta V100 GPUs with 32 GB GPU RAM accessed through MIT Supercloud~\cite{reutherInteractiveSupercomputing402018}. The time used for training depends on the training size and takes approximately 200 minutes for a training size of 960k images. The ensemble used for defect detection in this work consists of 4 networks and the training curve for these are  depicted in \textbf{Fig.~\ref{fig:training}}.

During training, the performance of the model is evaluated using a validation set. This set is generated in the same way as the training set. However, since this is synthetic data that comes from the same distribution as the training data it might not accurately represent the experimental images. As a result, improvements in performance on this validation set may sometimes be detrimental to the performance on experimental data.

To address this challenge we use several experimental images from previous experiments where we visually inspect network predictions, as described in the following section alongside the analysis. Specifically, we look for certain defective regions observed experimentally where a higher defect concentration should be present.

%

A clear challenge of our approach to training neural networks is getting a true sense of their performance, a challenge shared by any machine learning task trained on synthetic data. In future works we hope to address this using data-centric methods.

\subsection{4.7. Application of the Model} When we are satisfied with the performance of our model we can apply it on experimental data. In our case, we have trained our network to find the different defects, $V_{Br}$, $V_{Cr}$, $2V^{bi}_{Br}$ and $2V^{bi}_{Cr}$ in a 2L CrSBr. The work flow as a whole can be used with a variety of different targets such as atomic positions. Before applying the networks to the experimental images we apply pre-processing that matches the way we normalized the data during training, in this case normalizing the image by dividing by its maximum value.

\begin{figure*}[htp]
\scalebox{\figurescale}{\includegraphics[width=1\linewidth]{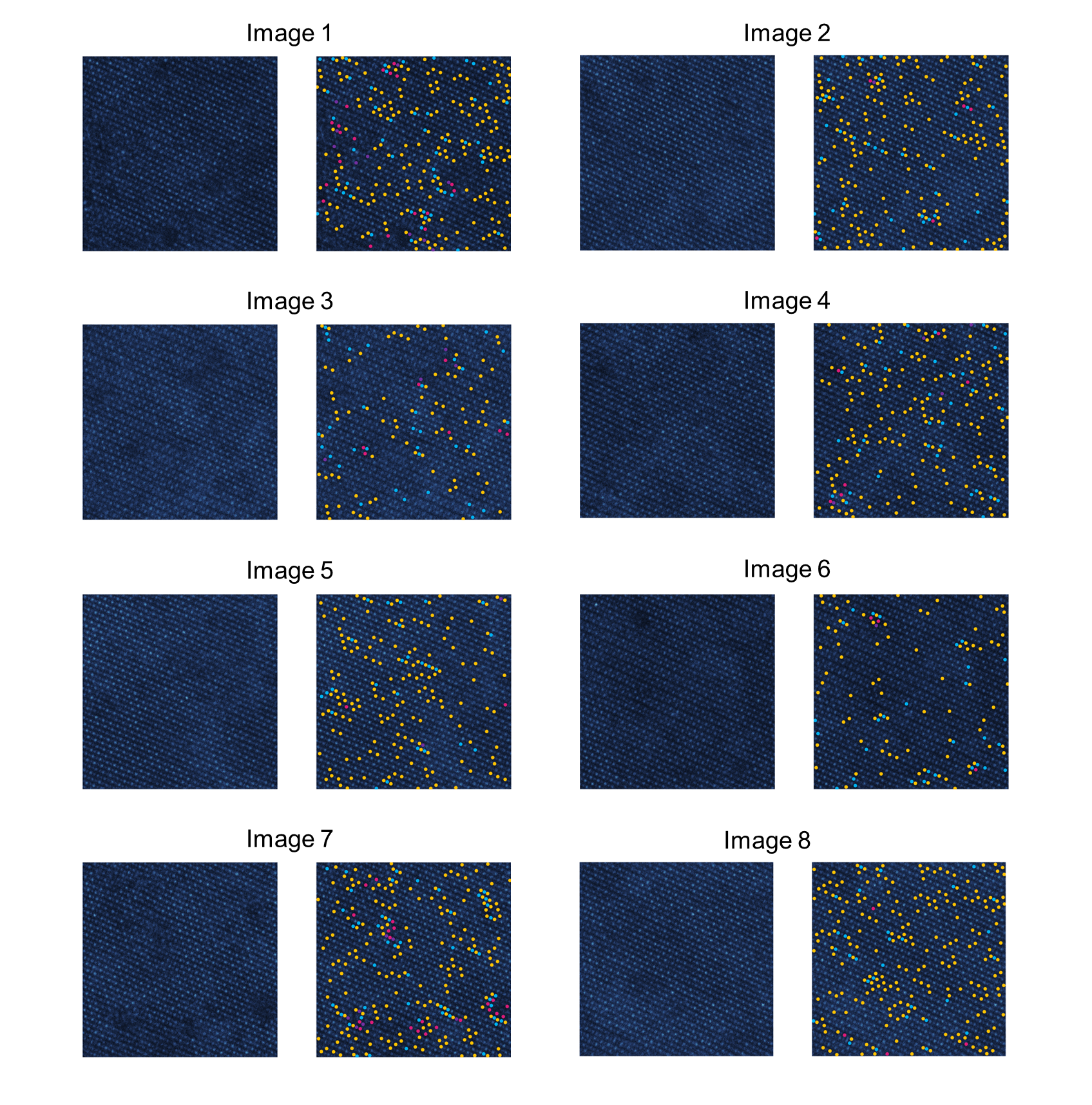}}
\caption{\label{fig:allimages}
\textbf{Application of DCNNs to experimental images.} 
  13 HAADF-STEM images and corresponding predictions using an ensemble of 4 DCNNs. The images depict different regions of the same 2L CrSBr flake and contain varying amounts of contamination, defective regions and therefore detected defects. A beam energy of 200keV and a beam current of 35pA was used. The total image dose is $9 \cdot 10^6 \text{e$^-$}\SI{}{\per\nano\meter\squared}$. Continued in \textbf{Fig.~\ref{fig:allimages2}}.
}
\end{figure*}

\begin{figure*}[htp]
\scalebox{\figurescale}{\includegraphics[width=1\linewidth]{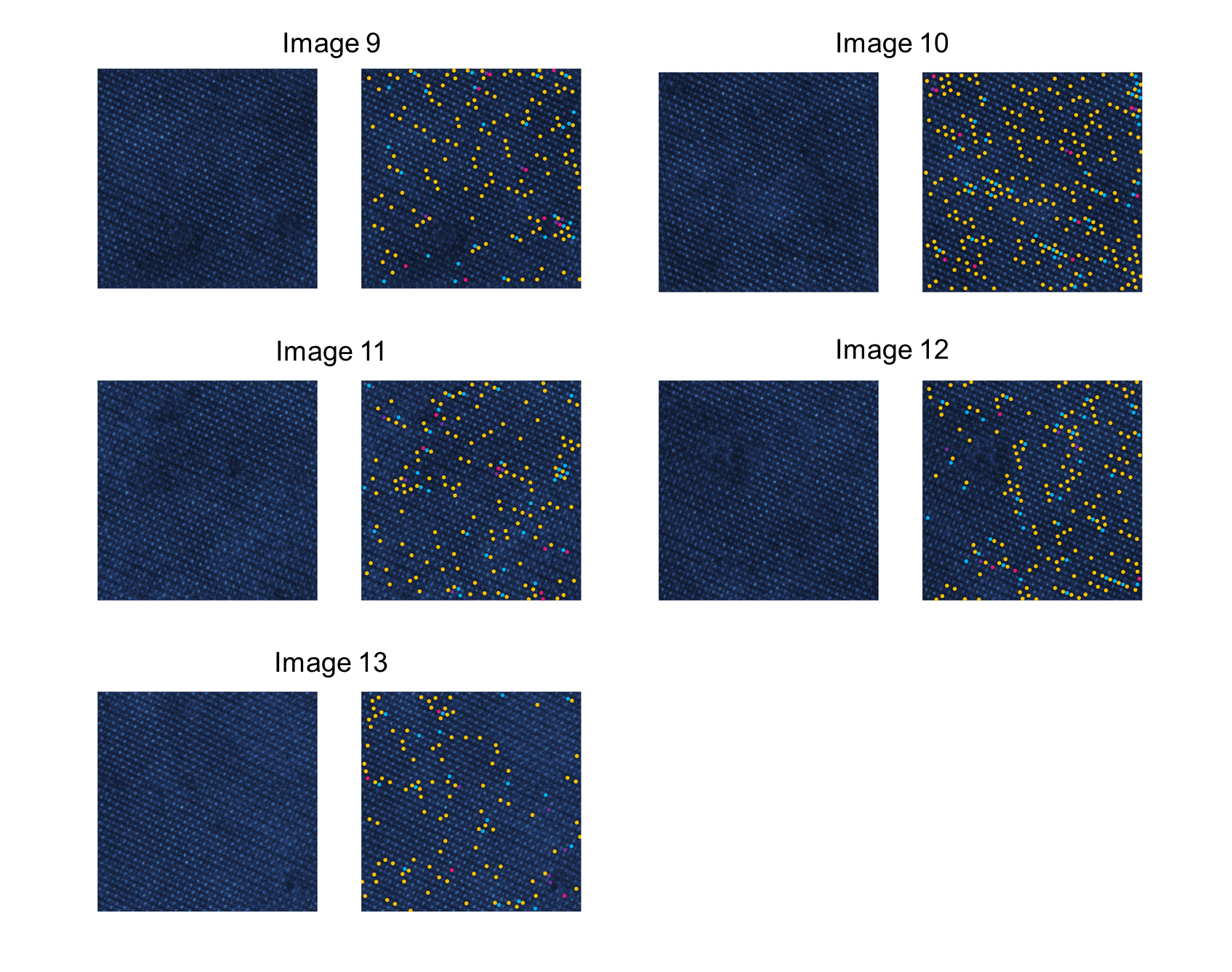}}
\caption{\label{fig:allimages2}
\textbf{Application of DCNNs to experimental images.} 
  Continued from \textbf{Fig.~\ref{fig:allimages}}.
}
\end{figure*}

\textbf{Fig.~\ref{fig:allimages}} and \textbf{Fig.~\ref{fig:allimages2}} show the detected defects alongside the original image for each of the 13 experimental HAADF-STEM images used in this manuscript. The different HAADF-STEM images were collected as described in the main text. The defects depicted are found using a threshold of 0.95 for $V_{Br}$ and 0.9 for all other defects.

\clearpage

\clearpage
\begin{figure*}[ht]
\scalebox{\figurescale}{\includegraphics[width=0.66\linewidth]{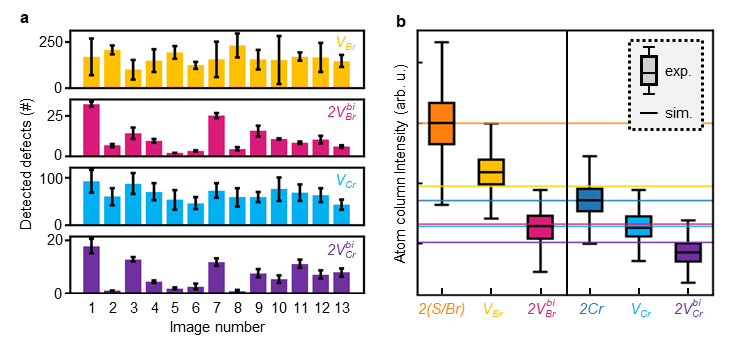}}
\caption{\label{fig:validation}
\textbf{Validation of DCNNs using experimental data.} 
  \textbf{a}, Benchmarking the ensemble of networks by comparing the amount of defects found by the individual networks. The errorbar corresponds to standard deviation of the number of defects found among networks. Variations in defects per image are due to larger defective regions in certain images (see \textbf{Fig.~\ref{fig:allimages}}).
   \textbf{b}, Atomic column intensity distribution for ten individual images. Distributions for defect free columns and columns with vacancies are shown and are normalized similar to \textbf{e}.
}
\end{figure*}

To cross-validate the consistency of our defect prediction using the ensemble, we also investigate the spread of the predictions within the ensemble where each DCNN was trained with identical parameters. This ensures that our DCNNs provide consistent predictions. We apply the ensemble to the set of 13 experimental HAADF-STEM images to obtain the absolute number of vacancy defects (see \textbf{Fig.~\ref{fig:validation}a}). We obtain consistent prediction results from the trained DCNN ensemble with variations in predictions that coincide with different degrees of damage in the respective experimental images, this is especially evident in the amount of $2V^{bi}_{Br}$ and $2V^{bi}_{Cr}$ detected which correlate. Furthermore, we also note that the spread in the detection of $V_{Br}$ is the largest as we expect given it has the lowest SNR compared to its pristine counterpart. We also note some images have imaging conditions and / or noise distributions that cause the networks to have worse performance, something we mitigate by using broad parameters during training.



Second, we investigate the S/Br and Cr atom column intensity distribution from the 13 experimental images and directly compare them to the atom column distribution of the different vacancy defect types (see \textbf{Fig.~\ref{fig:validation}b}). Here the intensity is calculated as the summation of pixels within a circular mask around the atom position with a radius of $\SI{30}{\pico\meter}$ determined to maximize contrast between Cr and S/Br columns. Based on the mean intensities of the distributions and simulated values for S/Br and Cr atom column and considering different vacancy defect configurations, we obtain excellent agreement. We also note the large spread in intensities even for the pristine Cr and S/Br columns caused by the challenging imaging conditions highlighting the need for methods such as machine learning.

Using these two outputs as additional checks, it is possible to evaluate the performance the different neural networks outside the validation set. While the evaluation is not perfect it serves as a helpful guide during parameter tuning. It however remains a key challenge since some conditions and areas will be out-of-distribution compared to the training data.

\begin{figure*}[ht]
\scalebox{\figurescale}{\includegraphics[width=0.66\linewidth]{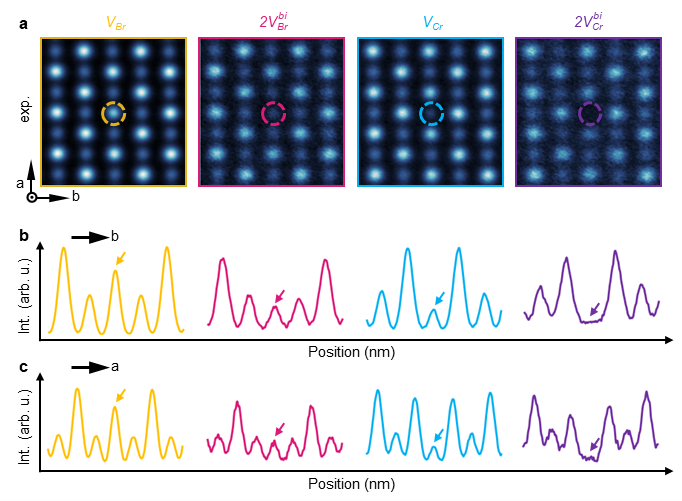}}
\caption{\label{fig:SI_classavg}
\textbf{Experimental class averages and line profiles along the b- and a-directions.} 
  \textbf{a}, Class-averages for each of the different vacancy defects we detect. Here shown with a Field of View of $1.2 \ \mathrm{nm}$.
   \textbf{b}, Line profile along the b-direction.
   \textbf{c}, Line profile along the a-direction.
}
\end{figure*}

\subsubsection{4.7.1. Class Averages from Experimental Images}

To investigate the different vacancies in CrSBr we construct class averages. To produce these class averages we adopt a two-step approach inspired by~\cite{ziatdinovBuildingExploringLibraries2019}. Here we first detect the different defects with our ensemble as described earlier. Afterwards we use another DCNN, the "atom spotter", to find the position of all the atomic columns, from which we can construct a lattice of column positions and types. The atom spotter is trained similarly to the defect detector networks, however, it is specialized in precisely determining the atomic position of atoms with few pm-precision. To utilize the benefit of this high precision we take every defect detected by the ensemble and find the associated site in the constructed lattice. This allows for much finer alignment of each of the defects when constructing the final class average. We apply this approach to the defects that were found in \textbf{Fig.~\ref{fig:allimages}} and \textbf{Fig.~\ref{fig:allimages2}} which results in the class-averages and the associated line profiles depicted in \textbf{Fig.~\ref{fig:SI_classavg}}. In 2L CrSBr, we obtain vacancy defect densities of $\SI{1.5}{} \cdot 10^{14}\SI{}{\per\centi\meter\squared}$ for $V_{Br}$, $\SI{3.2}{} \cdot 10^{13}\SI{}{\per\centi\meter\squared}$ for $V_{Cr}$, $\SI{7.7}{} \cdot 10^{12}\SI{}{\per\centi\meter\squared}$ for $2V^{bi}_{Br}$ and $\SI{4.3}{} \cdot 10^{12}\SI{}{\per\centi\meter\squared}$ for $2V^{bi}_{Cr}$. The densities are substantially higher as expected intrinsic defect concentrations~\cite{Klein.2022a} and we attribute them to the structural sensitivity of CrSBr, particularly when exposed to water or heat~\cite{Torres.2023} in combination with false positive assignments in these conditions.


\subsection{4.8. Benchmarking Different Convolutional Neural Network Architectures}

\begin{figure*}[ht]
\scalebox{\figurescale}{\includegraphics[width=1\linewidth]{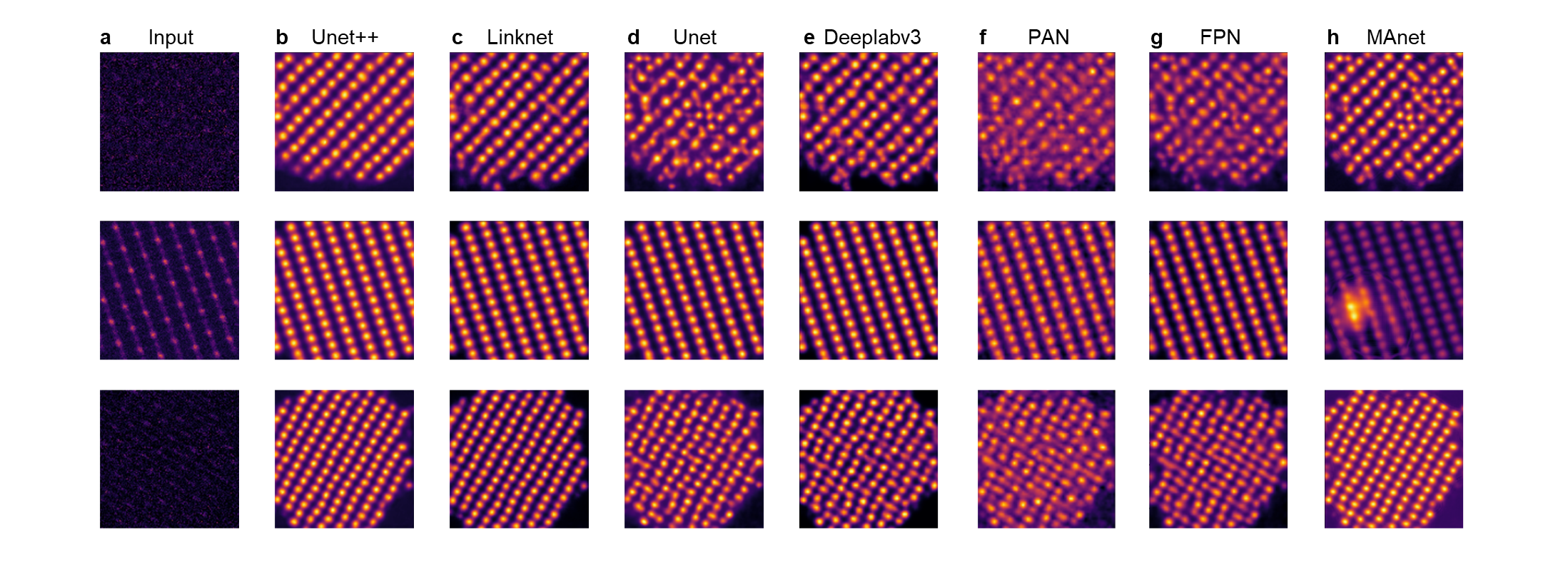}}
\caption{\label{fig:network_comparison}
\textbf{Benchmarking different neural network architectures.} 
In choosing the architecture to use for this manuscript we trained a collection of different network architectures to detect the positions of atomic columns in CrSBr. \textbf{a}, shows the input image. \textbf{b}, depicts the prediction when training a UNet++ network. \textbf{c}, LINKNET. \textbf{d}, UNet. \textbf{e}, Deeplab v3. \textbf{f}, PAN. \textbf{g}, FPN. \textbf{h}, MAnet. Each row depicts the prediction of the different models for simulated image matching the training conditions. Generally, we find UNet++ most suitable for our use-case.}
\end{figure*}

We visually compare 7 different segmentation architectures in \textbf{Fig.~\ref{fig:network_comparison}}. These models were trained on an atom position dataset of CrSBr and are available from~\cite{smp}. Transfer learning from ImageNet~\cite{imagenet} and equal training size was used for all networks. As evident the UNET++ architecture shows the most consistent predictions and captures the full details of the input. Specifically, it is able to detect the Cr columns even in challenging SNR conditions where the other networks often only find the SBr columns.

%
%

This benchmarking was done in the beginning of the project and for the task of finding atom positions. As such it might not represent the true performance of the networks that have been tuned for defect detection, but it serves as a helpful guide. 


\section{5. Experimental Images of Defect Complexes}
\begin{figure*}[h]
\scalebox{\figurescale}{\includegraphics[width=1\linewidth]{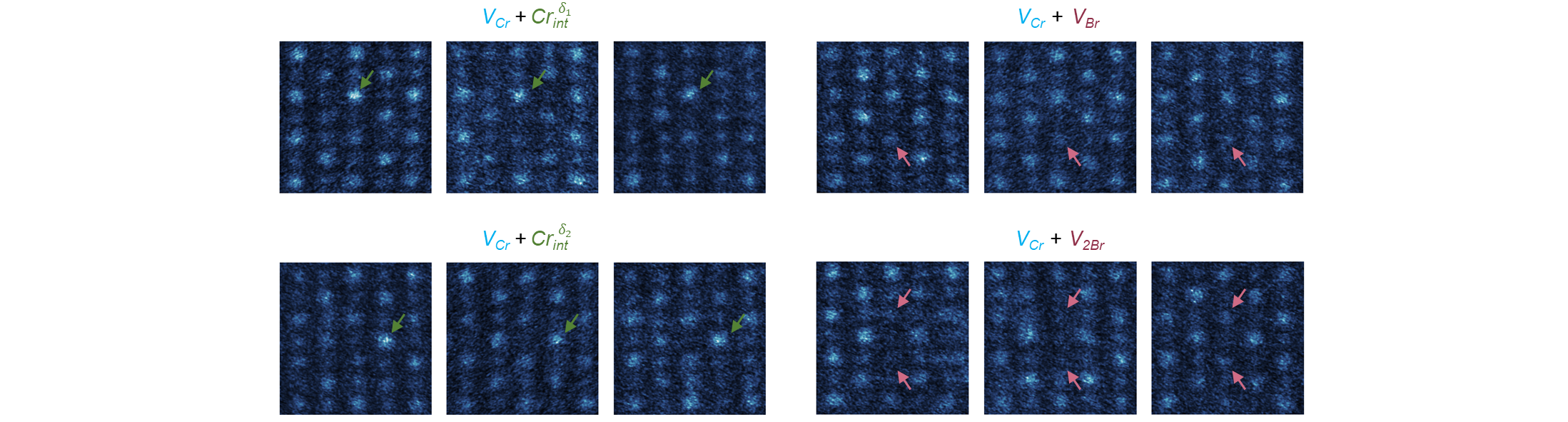}}
\caption{\label{fig:complexes_examples}
\textbf{HAADF-STEM images of defect complexes.}
Selection of manually selected defect complexes of the $V_{Cr}$+$Cr^{\delta 2}_{int}$, $V_{Cr}$+$Cr^{\delta 1}_{int}$, $V_{Cr}$+$V_{Br}$ and $V_{Cr}$+$V_{2Br}$.
}
\end{figure*}

Several examples of different defect complexes identified are shown in \textbf{Fig.~\ref{fig:complexes_examples}}. We note that even more complexes might be present in the compound.
%
%

\section{6. Surface and Van der Waals Interstitial Defects}

\begin{figure*}[ht]
\scalebox{\figurescale}{\includegraphics[width=1\linewidth]{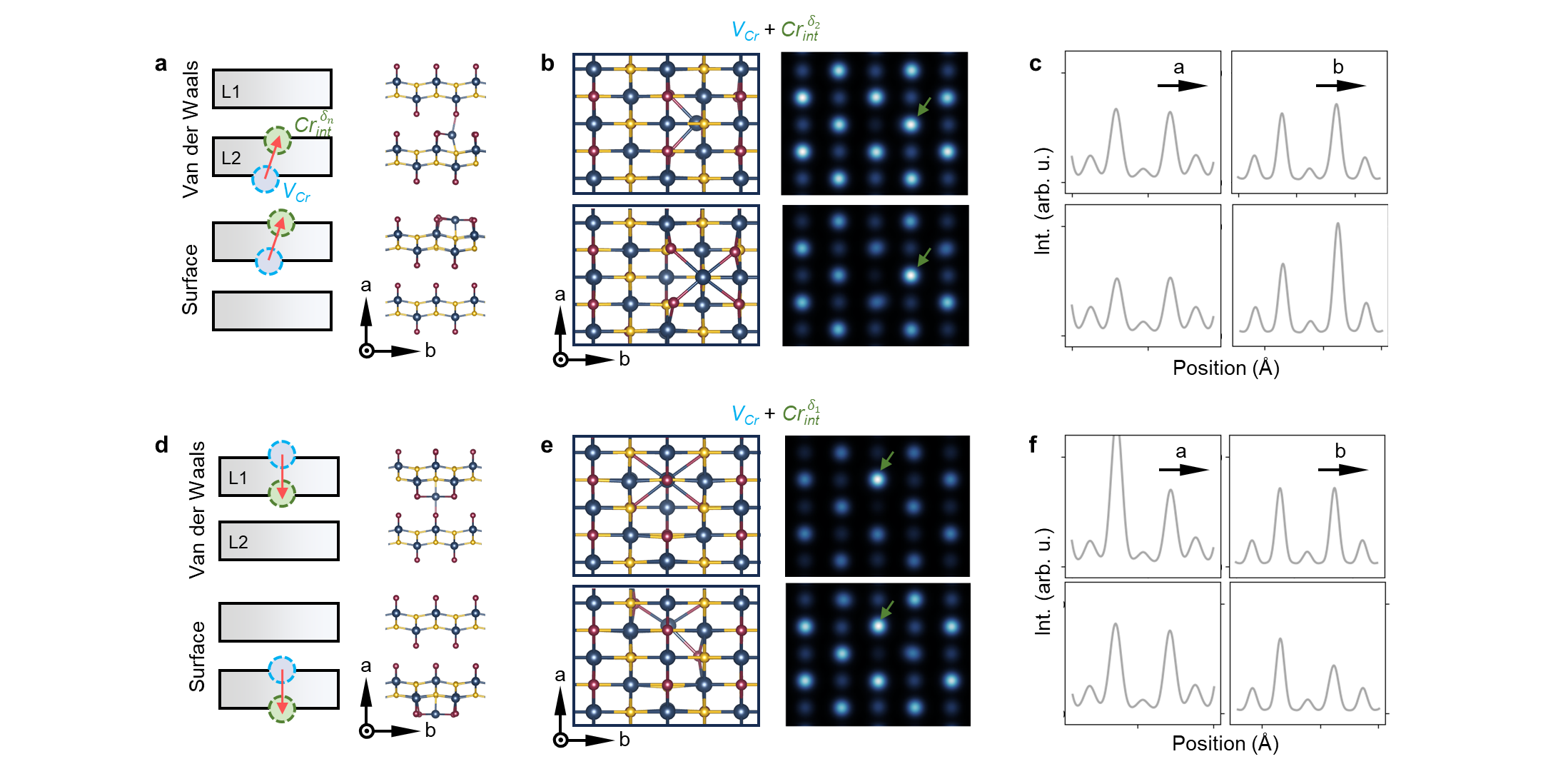}}
\caption{\label{fig:simulation_comparison}
\textbf{Surface and van der Waals interstitial vacancy defect complex simulation comparison.}
\textbf{a}, Schematic depiction of the $V_{Cr}$+$Cr^{\delta 2}_{int}$ interstitial vacancy defect complex and corresponding DFT relaxed crystal structure for a Cr interstitial situated either in the van der Waals gap or at the surface.
\textbf{b}, Top view of relaxed crystal structure and corresponding simulated HAADF-STEM image.
\textbf{c}, Line profiles along the a- and b-direction.
\textbf{d}, Schematic depiction of the $V_{Cr}$+$Cr^{\delta 1}_{int}$ interstitial vacancy defect complex and corresponding DFT relaxed crystal structure.
\textbf{e}, Top view of relaxed crystal structure and corresponding simulated HAADF-STEM image.
\textbf{f}, Line profiles along the a- and b-direction.
}
\end{figure*}

2L CrSBr exhibits two general types of interstitial defect sites, with the Cr interstitial atom either residing at the surface or in the van der Waals gap. We consider both types in our HAADF-STEM image simulations, as shown in \textbf{Fig.~\ref{fig:simulation_comparison}a-c} for the $V_{Cr}$+$Cr^{\delta 2}_{int}$ and in \textbf{Fig.~\ref{fig:simulation_comparison}d-f} for the $V_{Cr}$+$Cr^{\delta 1}_{int}$. We find that small offsets from the atom column position results in reduced contrast in HAADF-STEM. In our experiment, clearly resolved Cr interstitials may originate from either of the configurations shown in \textbf{Fig.~\ref{fig:simulation_comparison}}.

%
%

\section{7. Extended 1D Vacancy Line Defects}

\begin{figure*}[ht]
\scalebox{\figurescale}{\includegraphics[width=1\linewidth]{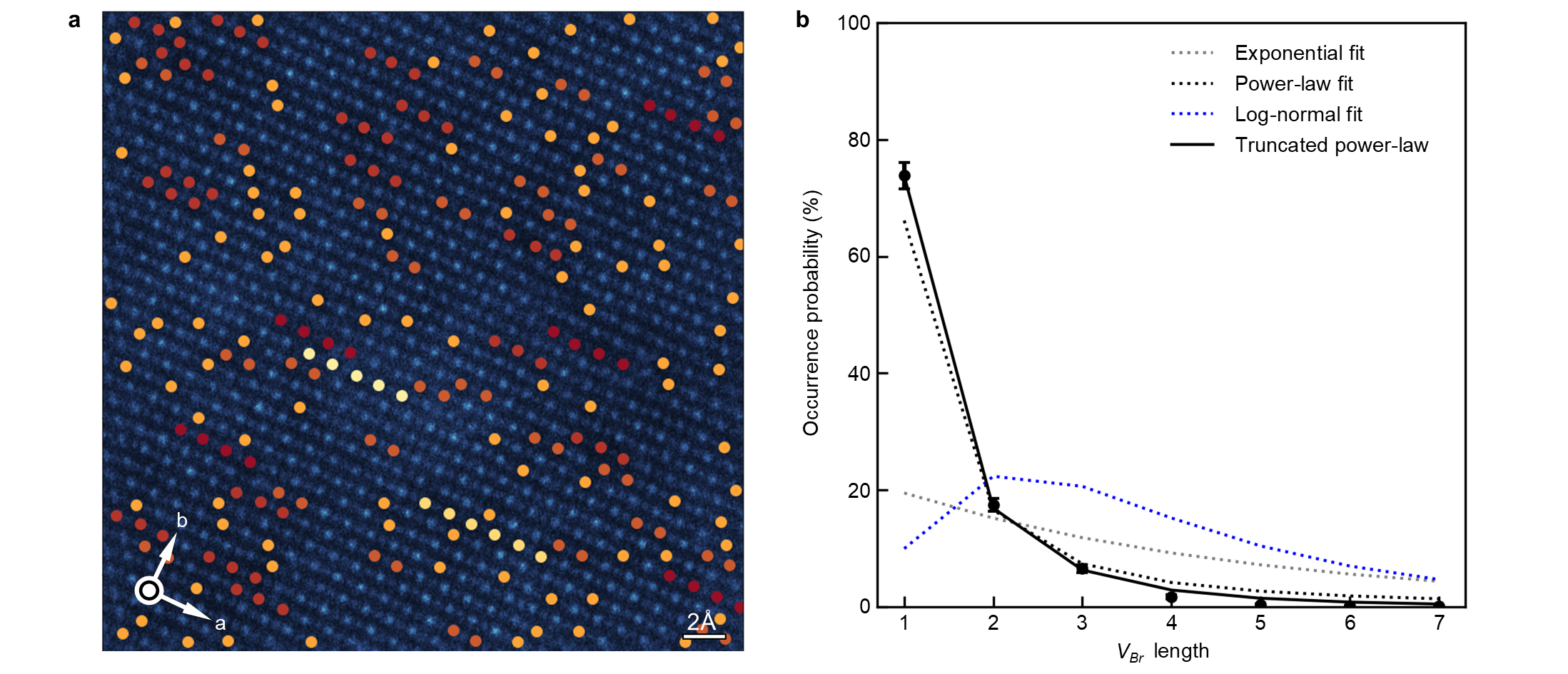}}
\caption{\label{fig:distribution}
\textbf{Statistical distribution of extended defect lines in 2L CrSBr.}
\textbf{a}, HAADF-STEM image with detected Br vacancy defect lines.
\textbf{b}, Occurrence probability of different Br vacancy lengths from 13 HAADF-STEM images (see \textbf{SM Figs.~\ref{fig:allimages}} and \textbf{~\ref{fig:allimages2}}) with four different types of fits. Best agreement is obtained using a (truncated) power-law distribution.
}
\end{figure*}

To determine the statistical distribution of Br vacancies, we count their spatial occurrence along the a-direction, particularly whether they appear as single vacancy or in extended vacancy lines. An exemplary HAADF-STEM image with overlaid Br vacancies, color-coded according to the length of the lines, is shown in \textbf{Fig.~\ref{fig:distribution}a}. We perform this process for all 13 images from \textbf{Fig.~\ref{fig:allimages}} and \textbf{Fig.~\ref{fig:allimages2}} and obtain the total count of each Br vacancy length. Using the total number of all Br vacancies as a reference, we calculate the probability of finding a defect line of a certain length by $P(x) = \frac{\text{count}(x)}{\sum_{x} (\text{count}(x) \cdot x)}$. This is shown in \textbf{Fig.~\ref{fig:distribution}b}. The distribution of defect lengths falls off sharply, with a slightly extended tail for longer defect lines. The same data on a double logarithmic scale is shown in \textbf{Fig.~4a} in the main manuscript.

To better understand the type of distribution that best describes the data and to interpret potential correlations in defect lengths, we fit different distributions, including an exponential distribution $P(x) = \lambda e^{-\lambda x}$, a log-normal distribution $P(x) = \frac{1}{x \sigma \sqrt{2\pi}} \exp \left( -\frac{(\ln x - \mu)^2}{2\sigma^2} \right)$, and a power-law distribution $P(x) = C x^{-\alpha}$. The latter shows a reasonable good agreement with the experimental distribution of Br defect lines, suggesting a correlated behavior in their occurrence. Nevertheless, we observe slight deviations, particularly at longer defect lengths. This is however expected, considering that the detection of Br vacancies using the DCNN is statistical in nature and can result in either false positives or false negatives. False positives can increase both individual $V_{Br}$ and longer extended lines while false negatives can increase shorter lines and underestimate longer lines. Moreover, the image field of view is limited in collected images which in turn can reduce the detected number of long defect lines. As a result, longer line defects are likely underestimated. For example, a 7-Br-vacancy-long defect may be counted as two shorter defects (e.g., 2 and 4) if one Br vacancy is missed, leading to an overestimation of shorter lengths.

To account for this, we also fit a truncated power law $P(x) = x^{-\alpha} e^{-x / x_{\text{max}}}$. The fit yields a truncation parameter of $\alpha = 3.29$, meaning that for defect lines longer than this, the probability of detection falls off more rapidly than expected. As mentioned, this is due to our conservative detection approach to avoid false positives, but other factors may also contribute, such as intrinsic material-related mechanisms, including internal stress limiting defect growth or potential self-repair mechanisms.

%
%

\begin{figure*}[ht]
\scalebox{\figurescale}{\includegraphics[width=1\linewidth]{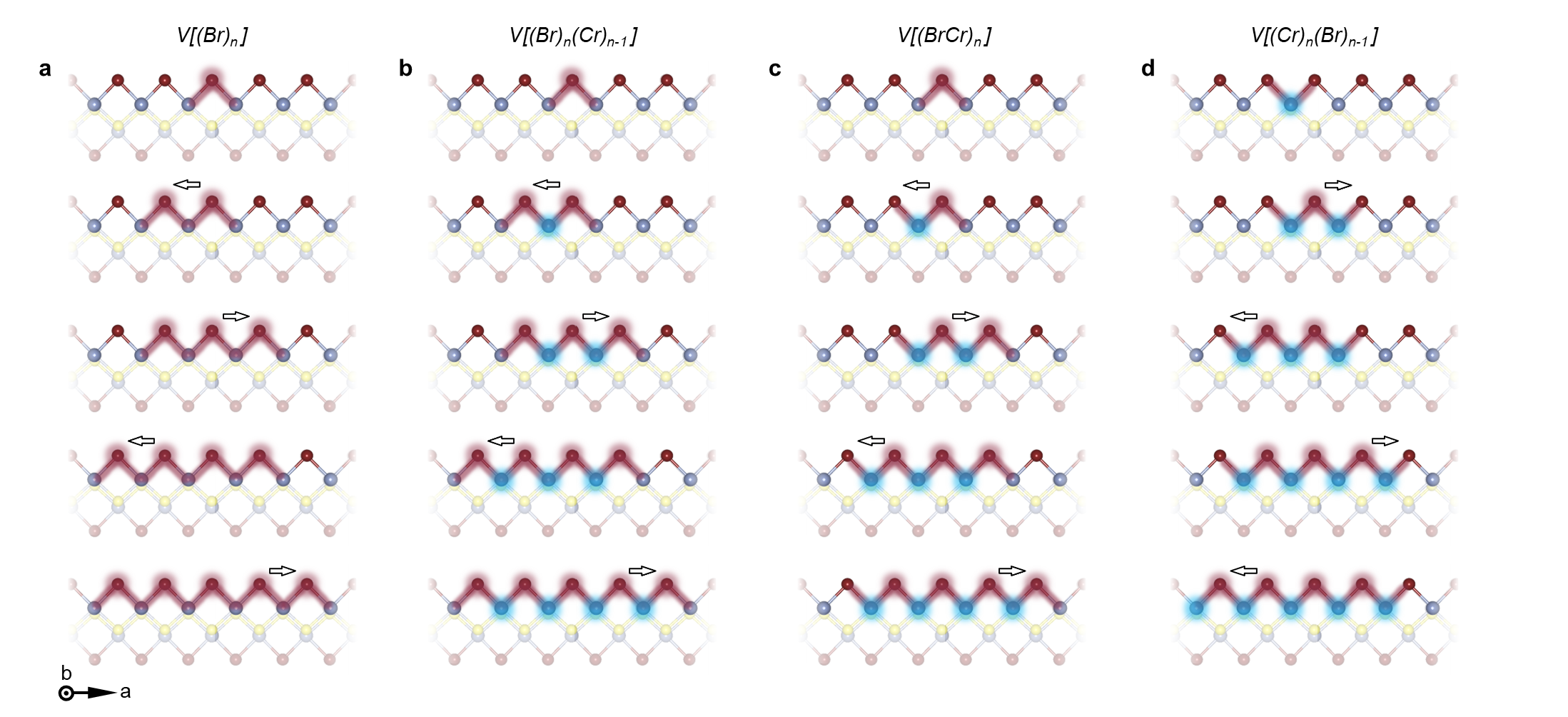}}
\caption{\label{fig:chains}
\textbf{Different types of 1D vacancy defect chains along the a-direction in CrSBr.}
\textbf{a}, Step by step formation of a line of Br vacancy defects $V[(Br)_n]$ and different combinations of Br vacancy and Cr vacancy defects \textbf{b}, $V[(Br)_n(Cr)_{n-1}]$ \textbf{c}, $V[(BrCr)_n]$ \textbf{d}, $V[(Cr)_n(Br)_{n-1}]$ with n defects.
}
\end{figure*}

For our 1D vacancy line defects, we consider four different combinations of vacancy defects and how they form step by step (see \textbf{Fig.~\ref{fig:chains}}). The most straightforward type of line defect is just a line of Br vacancy defects $V[(Br)_n]$ of length n (see \textbf{Fig.~\ref{fig:chains}a}). Other combinations include both Br and Cr vacancies (see \textbf{Fig.~\ref{fig:chains}b-d}).

%
%

\begin{figure*}[ht]
\scalebox{\figurescale}{\includegraphics[width=1\linewidth]{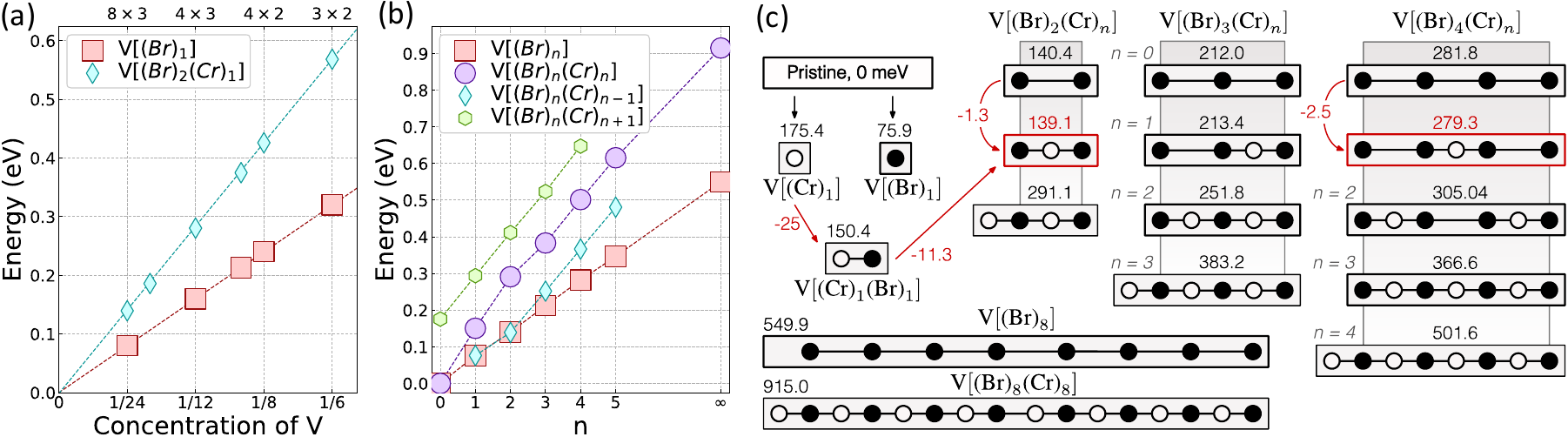}}
\caption{\label{fig:line_formation}
\textbf{Defect line formation energies.}
(a) Formation energy of a single V[(Br)$_1$] and V[(Br)$_2$(Cr)$_1$] defects as a function of concentration, expressed as $1/({N_x \times N_y})$, where $N_x$ and 
$N_y$ represent the dimensions of the supercell in the in-plane direction. 
A larger supercell corresponds to a lower concentration of defects.
(b) Formation energies for different types of 1D vacancy defect lines as a function of length n. The smaller the slope the more likely the type of defect line to form. A $8\times3\times1$ supercell was used.
(c) work flow illustrating the energetics of V[(Br)$_n$(Cr)$_m$] line vacancies in $8\times3\times1$ supercell. The numbers indicate formation energies (in meV). Red arrows highlight instances where the formation energy of line defects  is reduced by the presence of additional V[(Br)$_n$] or V[(Cr)$_m$] vacancies.
}
\end{figure*}

We utilize ab-initio calculations to investigate the formation energies of line defects, specifically examining how these energies vary with the defect length. To accommodate line defects of length \( 5a \) that are sufficiently spaced apart, we employ here a \( 8 \times 3 \times 1 \) supercell. This supercell is twice the size of the ones utilized \( 4 \times 3 \times 1 \) to study single defects, reducing the density of individual defects by half. Consequently, the corresponding formation energies are also halved, as illustrated in \textbf{Fig.~\ref{fig:line_formation}(a)} for the V[(Br)$_1$] and V[(Br)$_2$Cr] defects. In Figures \textbf{Fig.~\ref{fig:line_formation}(b)} and \textbf{(c)}, we present the formation energies of V[(Br)$_n$] line defects alongside varying concentrations of V[(Cr)$_m$] vacancies. The data show that pure V[(Br)$_n$] defects exhibit a lower formation energy when a small number of V[(Cr)$_m$] vacancies are added (specifically when \( n/m  \gtrsim 2 \)). For example, the defects V[(Br)$_2$Cr] and V[(Br)$_4$Cr] show slightly lower formation energies compared to their pure counterparts, V[(Br)$_2$] and V[(Br)$_4$], respectively. In contrast, all other combinations of V[(Br)$_n$] and V[(Cr)$_m$] vacancies are energetically less favorable.

In our experiment, particularly from the class averaged images of different lengths of vacancy lines, we observe not only $V_{Br}$ but also $V_{Cr}$. This suggests that the different defects combinations shown in \textbf{Fig.~\ref{fig:chains}} are to be expected to form under certain experimental conditions. Considering two surfaces in bilayer CrSBr, the observed vacancy line defects likely are formed at either of the surface.

%
%

\begin{figure*}[ht]
\scalebox{\figurescale}{\includegraphics[width=1\linewidth]{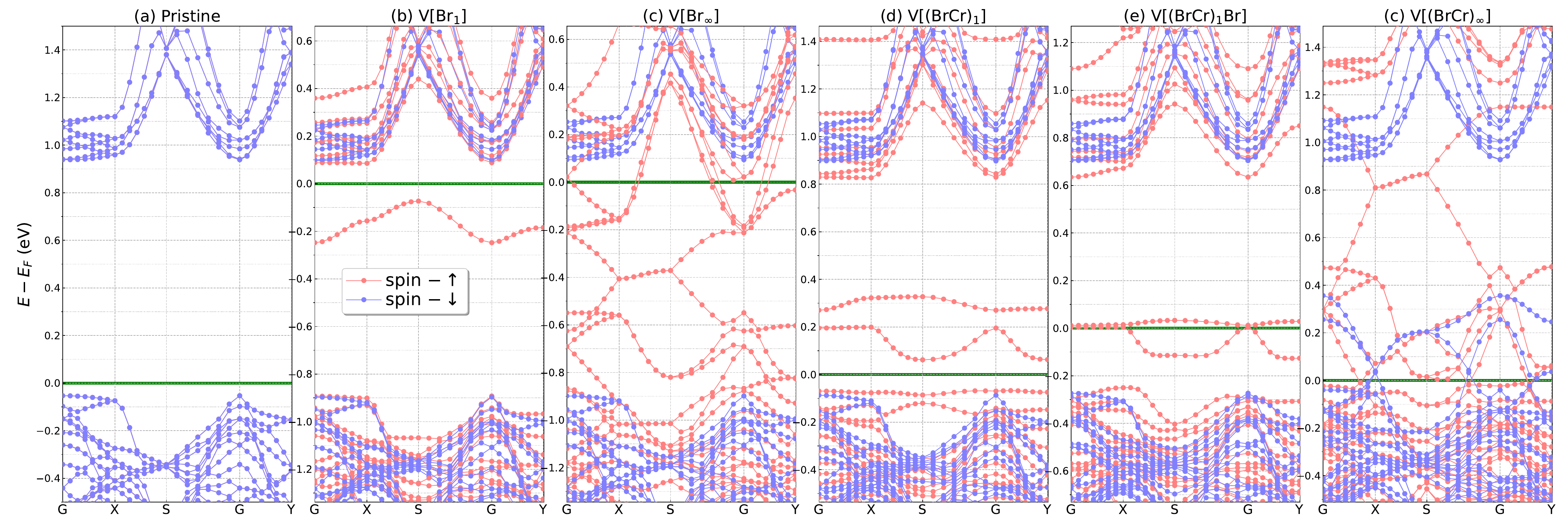}}
\caption{\label{fig:lines_electronic}
\textbf{Electronic structure of single defects and defect complexes in 2L CrSBr.}
Calculated band structure for pristine and defective 2L CrSBr in a $4\times3\times2$ supercell.
}
\end{figure*}

We continue calculating the electronic properties of the different variants of 1D vacancy defect lines (see \textbf{Fig.~\ref{fig:lines_electronic}}). To this end, we investigate the electronic band structure properties for (in)finite defect lines along the a-direction. We employ a $4\times3\times2$ supercell to minimize spurious interactions between replicas. For all cases discussed in \textbf{Fig.~\ref{fig:chains}}, we find a metallic behavior along the a- but also the b-direction. Moreover, the $V_{Br}$ forms a defect band that is energetically situated close the conduction band minimum expected to substantially contribute to the conductivity for the situation of n-doping~\cite{Wu.2022}. Furthermore, CrSBr is measured to be n-doped~\cite{Telford.2020}, likely due a high concentration of $V_{Br}$.

%
%


\section{8. Ab-Initio Calculations of Defect Complexes in CrSBr}

We utilize first-principle calculations to investigate the impact of (i) interstitial defects ($V_{Cr}$+$Cr^{\delta n}_{int}$) and (ii) vacancy complexes ($V_\text{X}$, X=Cr, CrS, CrBr, SBr) on the electronic properties of 2L CrSBr. 
To prevent structural distortions resulting from defect interactions, we employ a $4\times3\times1$ supercell in our computations to keep the defects at reasonable distances from one another. Given that experimental observations suggest that defects occur in either the top or bottom layer, albeit uncertain whether they are closer to the vacuum (surface) or the VdW gap (bulk), we analyze the effects of both defect locations on the electronic structures.

\begin{figure*}[ht]
\scalebox{\figurescale}{\includegraphics[width=1\linewidth]{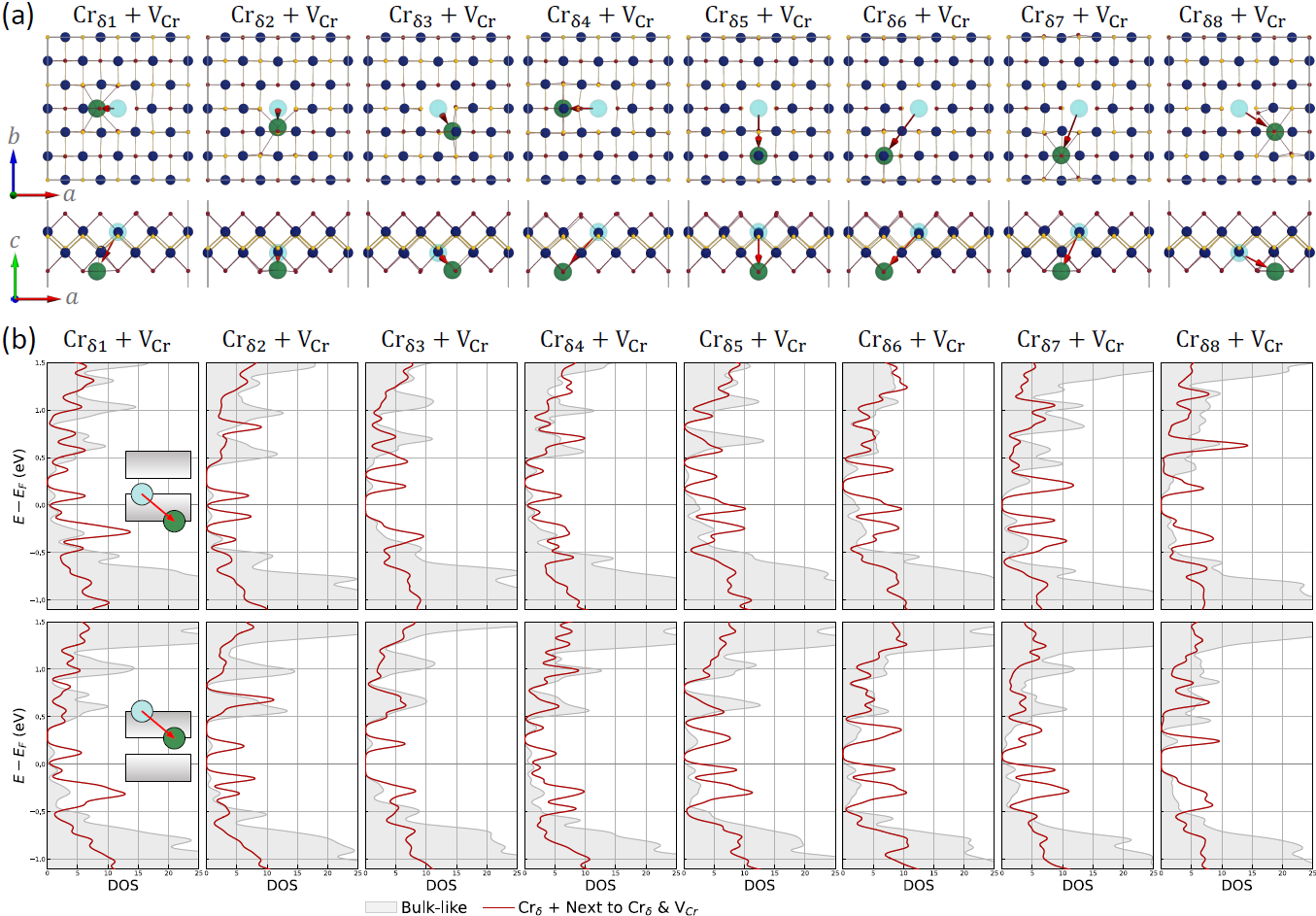}}
\caption{\label{SIfig:interstitials}
\textbf{Cr interstitial defects in 2L CrSBr.} 
(a) DFT-relaxed crystal structures of 8 distinct Cr interstitial defects ($Cr_\delta$) in 2L CrSBr.
(b) Density of States (DOSs) for the corresponding structures with $Cr_{\delta}$ in the vacuum (upper panels) and vdW layer (lower panels), respectively.
The lines in black and red denote bulk-like atoms and atoms associated with the impurities (Cr$_{\delta2}$ and all other atoms near $Cr_{\delta}$ and $V_\text{Cr}$).
%
}
\end{figure*}

\begin{figure*}[ht]
\scalebox{\figurescale}{\includegraphics[width=1\linewidth]{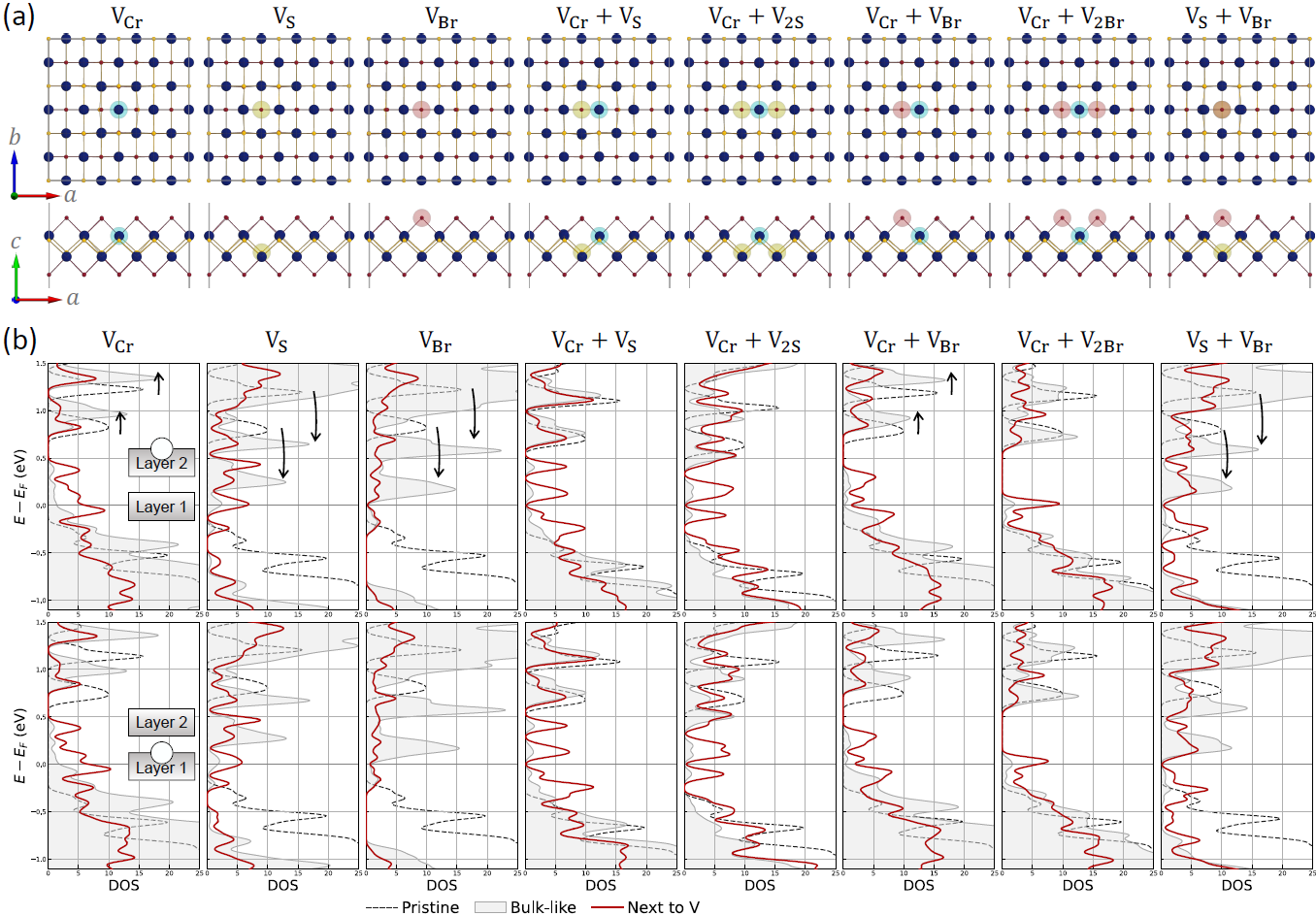}}
\caption{\label{SIfig:vac}
\textbf{Vacancy complexes in 2L CrSBr.} 
(a) DFT-relaxed crystal structures depicting seven distinct defect complexes in 2L CrSBr.
(b) Density of States (DOSs) for the corresponding structures showcasing vacancy complexes in the vacuum (upper panels) and VdW layer (lower panels). The dashed black lines represent the states of the pristine structure (aligned with states of atoms in the layer without vacancies). Gray and red lines indicate bulk-like atoms, and the atoms nearest to the vacancy complexes denoted as $V$.
}
\end{figure*}

\begin{figure*}[ht]
\scalebox{\figurescale}{\includegraphics[width=1\linewidth]{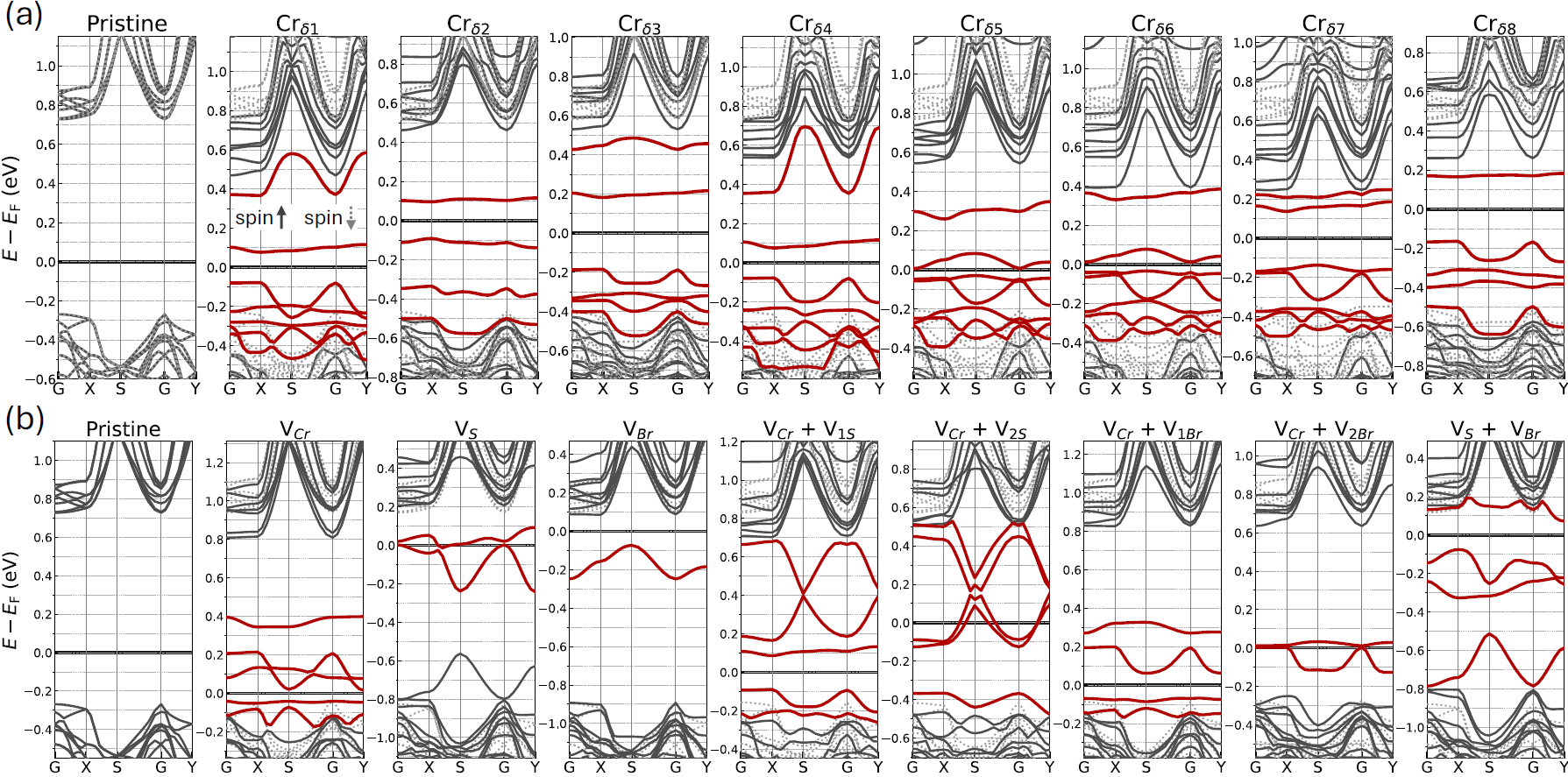}}
\caption{\label{SIfig:bands}
\textbf{Electronic structure of single defects and defect complexes in 2L CrSBr.} 
(a) interstitial defects V$_\text{Cr}$+Cr$^\delta$  and (b) vacancy complexes $V_\text{X}$, X=Cr, CrS, CrBr, SBr. Red lines indicated approximately defect-induced states. Solid and dotted lines differentiate the spin channels. The electronic structures displayed are for Cr$^\delta$ and vacancies positioned nearer to the vacuum than to the VdW region.
}
\end{figure*}

\begin{figure*}[ht]
\scalebox{\figurescale}{\includegraphics[width=1\linewidth]{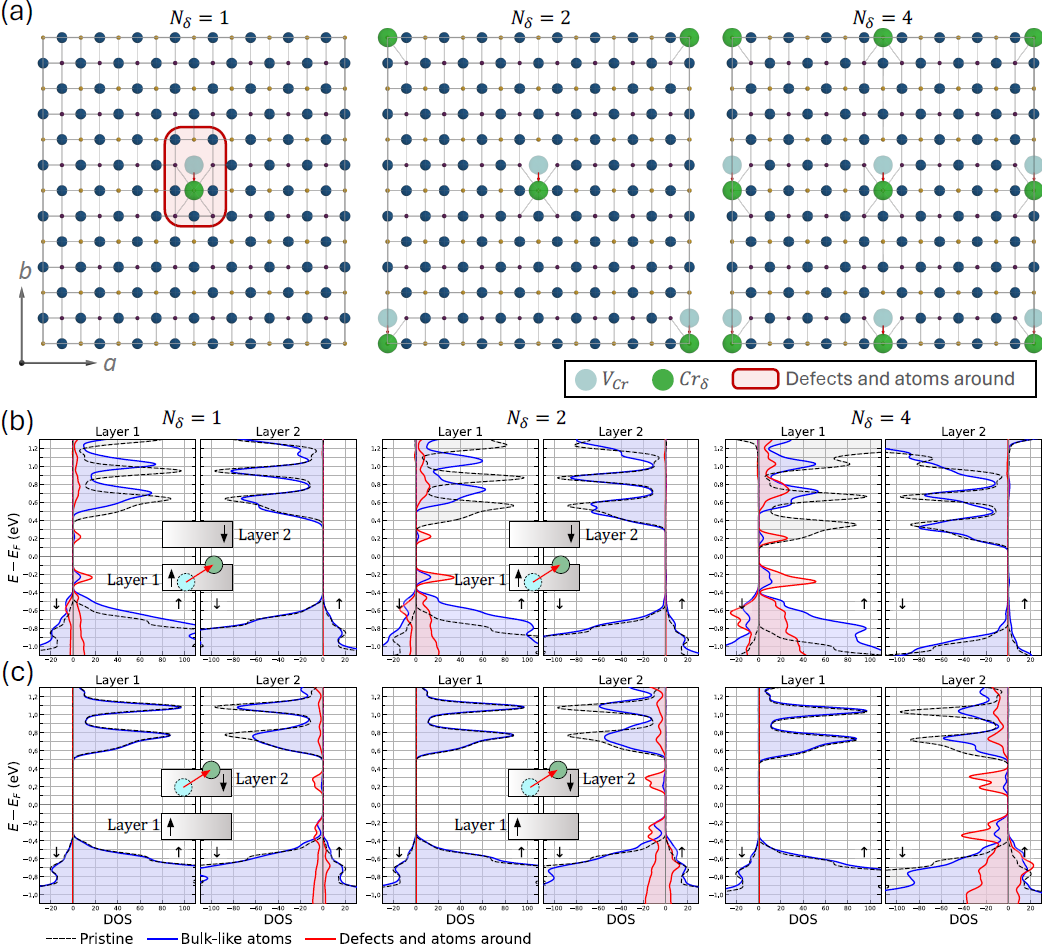}}
\caption{\label{SIfig:interstitial_density}
\textbf{Cr interstitial defects in 2L CrSBr.} 
Structures with varying concentrations of $Cr^{\delta 2}_{int}$ and $V_\text{Cr}$ defects in the $8\times6\times1$ bilayer supercell of CrSBr. In panels (b) and (c), the Density of States (DOS) is illustrated for $Cr^{\delta 2}_{int}$ positioned in the vdW layer and Vacuum, respectively. Positive and negative DOS values correspond to spin-up ($\uparrow$) and spin-down ($\downarrow$) states in layers 2 and 1, respectively.  
The black dotted line represents the DOS for the pristine structure, while the solid blue and red lines indicate bulk-like atoms and atoms connected to the impurities ($Cr^{\delta 2}_{int}$ and neighboring atoms near $Cr^{\delta 2}_{int}$ and $V_\text{Cr}$). 
The DOS plots for the pristine bilayer structure are aligned with the DOS for the layer devoid of defects in each corresponding bilayer structure containing $Cr^{\delta n}_{int}$ and $V_\text{Cr}$.
}
\end{figure*}

The relaxed atomic structures of vacancies, defect vacancy complexes and interstitial defect complexes are depicted in \textbf{Fig.~\ref{SIfig:interstitials}a} and \textbf{Fig.~\ref{SIfig:vac}a}. Please note that these figures represent atomic positions solely in the defected layer, with minor modifications in the atomic positions of the opposite layer due to weak vdW coupling. The corresponding density of states (DOS) for each defected structure are presented in \textbf{Fig.~\ref{SIfig:interstitials}b} and \textbf{Fig.~\ref{SIfig:vac}b}, with the upper and lower panels illustrating different defect locations in relation to the vacuum (surface) and vdW gap (bulk). The corresponding electronic band structures are presented in \textbf{Fig.~\ref{SIfig:bands}}. Here, we present the contributions to the DOS solely from the atoms in the distorted layer, whereas contributions from the undistorted layers remain consistent with those in the pristine structure, as illustrated in \textbf{SM Figs.~\ref{SIfig:interstitial_density}b, c} and discussed later in more detail.

From our analysis, we observe that it is a challenging task to separate the distinct contributions to the electronic states close to the Fermi level, delineating between the pure and bulk-related influences through atomic orbitals alone. Further comprehensive investigations employing localized (Wannier) orbitals are essential. This complexity arises because not all orbitals from atoms neighboring defects directly impact these states. Instead, it is a combination of selected orbitals, while others influence bulk states. Additionally, certain orbitals from bulk-like atoms in proximity to defects are influenced. Although their impact reduces rapidly with distance, their cumulative effect near the Fermi states is noticeable due to the larger number of such atoms compared to defects and neighboring atoms. It is important to note that the plotted DOS, which combines all atomic orbitals and atoms of a specific type, does not perfectly portray the intricate nature of the mixed states surrounding the Fermi level, serving more as an illustrative representation rather than an exact depiction.

A prevalent characteristic of the defected structures is that the states near the Fermi level primarily arise from the displaced $Cr^{\delta n}_{int}$ atoms, atoms surrounding both $Cr^{\delta n}_{int}$ and its vacancy $V_\text{Cr}$, and interstitial complexes $V_X$.
Conversely, these defects can either accumulate or deplete electrons on bulk-like atoms, leading to a shift in their DOS in comparison to the DOS of the pristine structure. This effect is demonstrated in \textbf{Fig.~\ref{SIfig:vac}b} and it depends on the density of the defects as it is elaborated upon in \textbf{Fig.~\ref{SIfig:interstitial_density}b} for further clarity and analysis. For instance, as shown in \textbf{Fig.~\ref{SIfig:interstitial_density}b}, the Cr (X = S or Br) vacancy can either attract or donate electrons from/to neighboring bulk-like atoms, causing their states to shift upwards/downwards in comparison to the pristine structure.

An important characteristic of the emerging states resulting from interstitial defects or vacancy complexes in most structures (except structures featuring only pure Br vacancy) is the co-existence of occupied and unoccupied states in proximity to the Fermi level, illustrated in \textbf{Fig.~\ref{SIfig:bands}}. This observation indicates that defect complexes within the 2L CrSBr have the potential to serve as quantum emitters resulting from intra-impurity excitations/relaxations.

Next, we investigate how the electronic properties vary with the density of interstitial defects. We focus on a configuration containing $\delta_2$ interstitial defects within an $8\times6\times1$ supercell, enabling the homogeneous distribution of 1, 2, and 4 ($N_\delta$) defects within this structure (see \textbf{Fig.~\ref{SIfig:interstitial_density}a}). Considering that each layer comprises 96 Cr atoms, the defect density for these scenarios can be represented as 1/96, 1/48, and 1/24 per Cr atom.  \textbf{Figures~\ref{SIfig:interstitial_density}b} and \textbf{c} display the DOSs when relocating Cr atoms from the vdW layer into the vacuum and vice versa. Each $N_\delta$ is further analyzed in two sub-panels, showcasing the DOS contributions from specific layers. Notably, the DOS from each layer, in the absence of defects, closely resembles that of the pristine structure. The states proximal to the Fermi level predominantly originate from the shifted Cr atoms, along with the atoms neighboring $Cr^{\delta n}_{int}$ and the Cr vacancy. Interestingly, the band gap between these Fermi-level states appears to remain relatively consistent across varying densities of Cr defects. A significant distinction is noted between 
$Cr^{\delta n}_{int}$ in the VdW layer 
(\textbf{Fig.~\ref{SIfig:interstitial_density}b}) and $Cr^{\delta n}_{int}$ in the vacuum 
(\textbf{Fig.~\ref{SIfig:interstitial_density}c}). In the former case, the contribution to the DOS from bulk-like atoms is heavily influenced by the concentration of $Cr^{\delta n}_{int}$, whereas the latter closely resembles the DOS of the pristine structure. This comparative analysis is crucial for understanding the impact of interstitial defects on the electronic properties.

\begin{figure*}[ht]
\scalebox{\figurescale}{\includegraphics[width=1\linewidth]{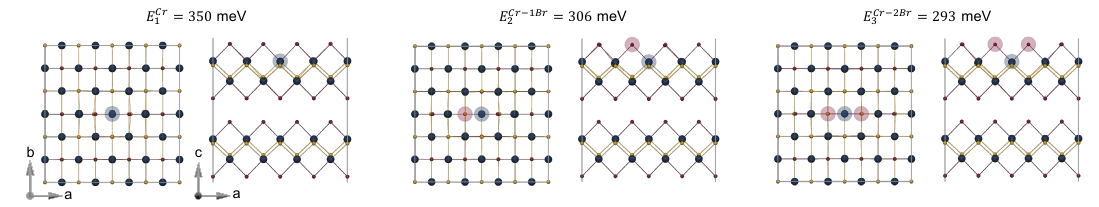}}
\caption{\label{SIfig:binding_Cr_Br_complex}
\textbf{Structural stability of 2L CrSBr in presence of vacancy defect complexes.} 
Relaxed defect structures and corresponding energy densities.
}
\end{figure*}

Concluding our exploration of the electronic properties of defects, we now turn our attention to the stability of 2L CrSBr in presence of these defect types. 
\textbf{Table.~\ref{tab_en_interstitial}} shows the energy densities needed to displace Cr atoms.
These values suggest that the displacement of Cr atoms is most energetically favorable when they are positioned not too close nor too far from their vacancies, specifically in the $\delta_2$ position. 
Moreover, the energy required to displace chromium atoms in all eight directions is lower when the displacement is towards the van der Waals region compared to displacement towards a vacuum. Interestingly, the energy required to displace a chromium atom from a layer where it is already displaced in the adjacent layer is lower compared to displacing the chromium atom individually.
To assess the energy cost involved in removing atoms from the 2L CrSBr, we calculate the binding energy, as detailed in the main text. 
\textbf{Table.~\ref{tab_en_vac}} and \textbf{Fig.~\ref{SIfig:binding_Cr_Br_complex}} illustrate the energies associated with the removal of different combinations of Cr, S, and Br atoms. Based on the data presented in \textbf{Table.~\ref{tab_en_vac}}, it can be concluded that the energetically most favorable option is removing Br atoms. A noteworthy observation is that removing Cr along with one or two of its closest Br atom neighbors, with which it shares strong ionic bonds, results in significantly lower energy costs compared to removing chromium atoms alone. Additionally, it is worth mentioning that atomic vacancies or their complexes incur lower energy expenses when they are closer to the van der Waals region rather than the vacuum.

%
%

%
%

{\renewcommand{\arraystretch}{1.5}
\begin{table}[t!]\centering
\begin{ruledtabular}
\begin{tabular}{l |r r r r r r r r}
 {Energy} &  $\delta_1$ & $\delta_2$ & $\delta_3$ & $\delta_4$ & $\delta_5$ & $\delta_6$ & $\delta_7$ & $\delta_8$   \\
 \hline
$E_\text{vac}$  &  170.715 & 118.728 & 139.516 & 154.823 & 189.391 & 189.368 & 188.547 & 154.580\\
$E_\text{VdW}$  &  149.230 & 105.442 & 126.123 & 133.833 & 167.061 & 166.273 & 173.046 & 146.051\\
$E_\text{vac+VdW}$  &  288.500 & 221.443 & 265.022 & 253.482 & 347.384 & 348.113 & 356.431 & 295.801\\ \hline
$\Delta E_\text{vac+VdW}$  & -31.446 & -2.727 & -0.617 & -35.175 & -9.068 & -7.528 & -5.162 & -4.830 \\
%
\end{tabular}
\caption{
\textbf{Structural stability of 2L CrSBr in presence of interstitial defects.}
Energy densities of 8 distinct Cr interstitial defects V$_\text{Cr}$+Cr$^\delta$ in 2L CrSBr. The energies are provided relative to the pristine structure (in meV units) and are divided by 
$N_x \times N_y \times N_z$, where $N_x=4$, $N_y=3$, and $N_z=2$ are the number of unit cells in the three Cartesian directions.
The energies $E_\text{vac}$, $E_\text{VdW}$, and $E_{12}$ correspond to structures with Cr$^\delta$ defects situated in the vacuum region, the Van der Waals region, and both regions, respectively. The final row displays the increase in energy density resulting from the displacement of Cr$^\delta$ atoms in both regions, $\Delta E_\text{vac+VdW} = E_\text{vac+VdW} - (E_\text{vac} + E_\text{VdW})$ (contrasted with the energy density of Cr$^\delta$ defects in the individual regions).
}
\label{tab_en_interstitial}
\end{ruledtabular}
\end{table}}

{\renewcommand{\arraystretch}{1.5}
\begin{table}[t!]\centering
\begin{ruledtabular}
\begin{tabular}{l |r r r r r r r r}
 {Energy} & Cr & S & Br & Cr+1S & Cr+2S & Cr+1Br & Cr+2Br & SBr\\
 \hline
$E_\text{vac}$ &  349.769 & 277.950 & 165.139 & 602.238 & 707.412 & 306.245 & 293.404 & 407.330 \\
$E_\text{VdW}$ &  350.472 & 278.956 & 161.090 & 602.667 & 710.300 & 300.804 & 281.899 & 402.390 \\
$E_\text{vac+VdW}$ &  699.147 & 556.908 & 325.429 & 1203.677 & 1415.984 & 606.754 & 575.254 & 809.323 \\\hline
$\Delta E_\text{vac+VdW}$ &  -1.093 &  0.002 & -0.800 & -1.228 & -1.728 & -0.295 & -0.048 & -0.398 \\
%
\end{tabular}
\caption{
\textbf{Structural stability of 2L CrSBr in presence of vacancy defects.}
Binding energy density for specific types of atomic vacancies and their complexes. The energies $E_\text{vac}$, $E_\text{VdW}$, and $E_\text{vac+VdW}$ correspond to structures vacancies situated closer to the vacuum region, the Van der Waals region, and both regions, respectively. 
The expression for the binding energy density is $E = [(E_{N-n} + E_n) - E_N]/(N_x \times N_y \times N_z)$, where $E_N$ is the energy of the pristine structure made of $N$ atoms, $E_n$ is the energy of $n$ atoms removed from the pristine structure, and $E_{N-n}$ is the energy of the structure with $n$ vacancy atoms.
Here, $N_x$, $N_y$, and $N_z$ are the number of unit cells in the three Cartesian directions.
The final row displays the increase in binding energy density resulting from the formation of the same vacancies in both regions, $\Delta E_\text{vac+VdW} = E_\text{vac+VdW} - (E_\text{vac} + E_\text{VdW})$.
}
\label{tab_en_vac}
\end{ruledtabular}
\end{table}}

{\renewcommand{\arraystretch}{1.}
\begin{table}[t!]\centering
\begin{tabular}{p{2.5cm}|p{2.cm}|p{2.5cm}|p{2.5cm}|p{2.cm}|p{2.cm}}
\toprule
\multirow{2}{*}{Structure} & {Orientation of} & \multirow{2}{*}{Type}  &\multirow{2}{*}{Crystal system} & 
\multicolumn{2}{c}{Point group} \\ \cline{5-6}
 & the defect & & & International  & Schoenflies   \\\toprule
 Pristine &  & & Orthorhombic& mm2 & $C_{2v}$\\ \hline 
 $V_{X}$ &  & Cr, Br, S, SBr &  Orthorhombic  & mm2 & {$C_{2v}$}  \\ \hline 
 \multirow{2}{*}{$V_{X}+V_{1Y}$} 
&$(\parallel x) \vee (\parallel y) $      & Cr-1S, Cr-1Br  & Monoclinic  & m & {$C_S$} \\ 
&$(\nparallel x) \wedge ( \nparallel y) $ & not studied    & Triclinic   & 1 & $C_1$   \\ \hline 
 \multirow{2}{*}{$V_{X}+V_{2Y}$} 
& $(\parallel x) \vee (\parallel y) $    & Cr-2S, Cr-2Br & Orthorhombic  & mm2 &  $C_{2v}$  \\ 
&$(\nparallel x) \wedge (\nparallel y) $ & not studied   & Monoclinic    & 2   & $C_2$     \\ \hline 
\multirow{2}{*}{$V_\text{Cr}+\text{Cr}^\delta$} &  
$( \parallel x) \vee (\parallel y) $ & $\delta_1$, $\delta_2$, $\delta_4$, and $\delta_5$ & Monoclinic  & m & {$C_S$} \\ 
&$(\nparallel x) \wedge (\nparallel y) $  &$\delta_3$, $\delta_6$, $\delta_7$, and $\delta_8$& Triclinic  & 1 & {\bf $C_1$} \\ \toprule
\end{tabular}
\caption{
\textbf{Symmetries of 2L CrSBr in presence of defects.}
}
\label{tab_sym}
\end{table}}


\FloatBarrier
\centering
\textbf{References}
\bibliographystyle{apsrev}
\bibliography{full}